\documentclass[%
 reprint,
 amsmath,amssymb,aps,
longbibliography]{revtex4-2}

\usepackage{graphicx}
\usepackage{dcolumn}
\usepackage{bm}

\usepackage{enumitem}
\usepackage{soul}
\usepackage{ulem}
\usepackage{xcolor}
\usepackage{multirow}
\usepackage{url}

\begin{document}

\preprint{APS}

\title{Characterization and optimized engineering of bosonic quantum interfaces under single-mode operational constraints}

\author{Pak-Tik Fong$^1$}
\author{Sheung Chi Poon$^2$}%
\author{Hoi-Kwan Lau$^{1,3}$}
\affiliation{%
$^1$Department of Physics, Simon Fraser University, Burnaby, British Columbia V5A 1S6, Canada\\
$^2$Department of Physics, The Chinese University of Hong Kong, Shatin, Hong Kong, China\\
$^3$Quantum Algorithms Institute, Surrey, British Columbia V3T 5X3, Canada}


\date{\today}

\begin{abstract}
Controlling the quantum interface between two bosonic modes is essential in countless implementations of quantum information processing. However, full controllability is rarely achieved in most platforms due to specific physical limitations. In this work, we completely characterize the linear two-mode interfaces under the most pessimistic restriction that only single-mode operation is available. When arbitrary Gaussian single-mode operations can be applied to both modes, we find that every interface can be characterized by an invariant transmission strength. Moreover, in the practical situation that squeezing is restricted in one of the modes, we discover two additional quantities, irreducible squeezing and irreducible shearing, that are invariant under the allowable controls. By using this characterization, we develop systematic strategies to engineer an arbitrary linear interface through cascading multiple fixed component interfaces. Without squeezing restriction, our protocol is optimal and requires at most three component interfaces. Under the squeezing constraint, our protocol can be extended to engineer also the additional invariants by using no more than two more rounds of cascade. We also propose the remote squeezing scheme to tackle the squeezing restriction through interfacing with an active auxiliary mode.
\end{abstract}

\maketitle


\section{Introduction}

Bosonic quantum systems are ubiquitous in implementing platforms of quantum technologies. For example, photon is widely used as an information carrier in quantum communication systems \cite{quantum_com2007} and photonic quantum computers \cite{KML2001,RMP2007}. Superconducting microwave resonator is a key element in the circuit quantum electrodynamics quantum computing architecture \cite{superconducting2004a,superconducting2004b}, bosonic coding implemented in this platform has achieved break-even point of quantum error correction \cite{superconducting2016}. The controllable phonon-photon interaction in optomechanical systems \cite{cavityopto2014} has been utilized to implement cutting-edge technologies \cite{optpmechanics_technologies2022}, such as microwave-to-optical transduction \cite{signal_tran_2012}, topologically protected transport \cite{topo2018,topo2022} and small-scale thermal machines \cite{thermo2014a,thermo2015}.
In the low excitation regime, an ensemble of spin also behaves as bosonic \cite{hp_transformation1940}; such platform has been proposed to be a robust quantum memory for light \cite{atomic_ensemble1,atomic_ensemble2, atomic_ensemble3} and microwave \cite{spin_ensemble_register, spin_ensemble_memory, spin_ensemble_memory2}, as well as the building block of quantum repeaters \cite{DLCZ, atom_quantum_repeater2011}. Generally, bosonic systems can also be applied in continuous variable (CV) quantum computing \cite{SethLloyd1998,rmp2005,rmp2012}, quantum sensing and metrology \cite{ BSJ1986, SethLloyd2006, Jonathan2010}, demonstrating quantum computational advantages \cite{BSD2011,BSD2013,BSD2013_1,BSD2013_2} and quantum chemistry simulation \cite{molecular_vibronic_spectra2015, sim_vibration_photonics2018}.

To realize the above implementations and applications, it is essential to generate a Gaussian interface with purpose-specific strength and type. Gaussian interfaces are categorized into seven types in Ref.~\cite{lau2019} and each type is utilized with specific purposes. They are Beam-splitter (BS), two-mode squeezing (TMS), quantum non-demolition interface (QNDI), SWAP, Identity interface, swapped-TMS (sTMS) and swapped-QNDI (sQNDI). BS type interfaces between modes are required for boson sampling \cite{BSD2013_exp,BSD2020} and quantum walk simulators \cite{quantumwalk1999,quantumwalk2003}. Two-mode-squeezing (TMS) is the interface underlying spontaneous parametric down-conversion photon generation \cite{SPD1,SPD2,SPD3} and phase-insensitive amplification \cite{caves1982}. Quantum non-demolition interface (QNDI) is utilized in generating CV cluster states for measurement-based quantum computing \cite{Nicolas2006,Gu2009} and quantum secret sharing \cite{Lau:2013wb, kogias2017unconditional, zhou2018quantum}. There has been increasing interest in engineering SWAP for transducing information \cite{atomic_ensemble3, transducer2014, MOtransduction2014, MOtransduction2018} between different components of hybrid quantum systems \cite{Kurizki2015, Clerk2020}, and rapidly cooling trapped ions \cite{lau2014diabatic, sagesser2020robust}. Identity interfaces are sometimes engineered to avoid unnecessary excitation, such as in heatingless ion separation and merging \cite{lau2012proposal, sutherland2021motional}. sTMS and sQNDI are newly discovered in Ref.~\cite{lau2019}. sQNDI allows quantum transduction with less stringent conditions \cite{lau2019, Jiang2022}, while there is no proposed application of sTMS yet.  

Two-mode linear interfaces have been widely studied by different scientific communities under different names. For examples, in quantum information processing they are usually used as Gaussian logic gates in CV quantum computing, or a description of Gaussian quantum channels in the open quantum system \cite{rmp2012}. In condensed matter physics, it is commonly known as Bogoliubov transformation \cite{bogoliubov, bogoliubov2}. In mathematics, the transformation introduced by such interface is studied under the name of symplectic transformation \cite{sympletic_book}. In this work, since we focus on the processes' properties in interacting two independent bosonic degrees of freedom,  we will consistently denote the process as "interface".

Implementing a two-mode interface with the desired type and strength is generally challenging due to various experimental limitations. First the type of interaction is usually restricted by the physical properties of the platform. For examples, the Stokes and anti-Stokes interactions in electro-optical \cite{optomechanical_tsang2011} or optomechanical systems \cite{cavityopto2014} typically generate either a TMS or BS interface; QNDI arises naturally between light and atoms ensemble as the interaction is bilinear in the quadrature of each system \cite{atomic_ensemble3,Filip2010}. Furthermore, there are limitations in the interaction strength. For example, 
the strength of the quantum light-atom interface is characterized by the optical depth of the atomic ensemble \cite{atom_light_ensemble_rmp2010} that cannot be arbitrarily large, since the dense atomic ensemble is avoided to prevent unwanted interaction, such as the dephasing induced by the dipole interaction \cite{atomdephasing2021}.

One might naively think that the unwanted properties of an imperfect interface can be removed by applying single-mode operations. However, this idea is disproved because the interfaces with different characteristic properties are not inter-convertible by single-mode operations \cite{lau2019}. To overcome this fundamental limitation, one approach is to cascade multiple rounds of the restricted interfaces. Earlier proposals suggested that SWAP can be implemented by cascading multiple QNDIs \cite{Filip2010,Warwick2016,Filip2017,Filip2018}. Recently,
Lau and Clerk \cite{lau2019} developed a more general scheme to construct SWAP and sQND cascading up to any six fixed interfaces. 
Zhang \textit{et al.} \cite{Jiang2021} generalized this protocol to implement arbitrary Gaussian operations by applying exponentially many rounds of the same multi-mode interface. In spite of the possibilities introduced by these protocols, two important practical issues have not been fully addressed:
\begin{enumerate}
    \item Optimality in rounds of applied interface: all the above protocols have not explored the minimum number of the required interfaces. It is generally beneficial to apply fewer rounds of interface because engineering inter-mode interaction is challenging, especially for the hybrid system.
    \item Squeezing: some existing protocols assume that all modes can be squeezed, so they cannot be applied to the platforms involving systems that squeezing is challenging, e.g. optical and spin-ensemble systems.
\end{enumerate}

Our work will address both of these issues. In the first half, we focus on the situation every single-mode operation is allowed. We find that every interface can be characterized by its invariant transmission strength, and every interface sharing the same transmission strength is inter-convertible. We then develop the protocols for engineering an interface with arbitrary transmission strength. By applying single-mode controls between at most three directly available interfaces with predetermined strength, referred to as component interfaces hereafter, our protocol can manipulate the overall transmission strength through amplification and interference. We prove that our protocol is optimal in the sense that it involves the fewest possible number of component interfaces.

In the second half, we extend the study to the restricted scenario that squeezing is available to only one of the modes. We discover two more quantities, irreducible squeezing and irreducible shearing, that are invariant under this additional operational restriction. To engineer an interface with any desired magnitudes of these invariant quantities, we develop a protocol that involves at most four component interfaces. To resolve the squeezing restriction, we also introduce the remote squeezing scheme that squeezes the restricted mode by interfacing it with an active auxiliary mode. 

Our paper is organized as follows. We start with the general scenario and review the interfaces' characterization by the transmission strength in Sec.~\ref{classification_begin}. Then, we present the optimal interface engineering protocols in Sec.~\ref{sec_setup}. In Sec.~\ref{sit_with_squeezing_restriction}, we discuss the new invariant parameters for characterizing an interface in the squeezing-restricted scenario. In Sec.~\ref{modified_protocol}, we propose the squeezing-restricted protocols that construct interfaces with any desired characteristic parameters.
Particularly, we present the remote squeezing protocol that can serve as a new method to overcome the squeezing restriction in Sec.~\ref{remote_gate}.Sec.~\ref{conclusion} concludes the paper. 

\section{Characterization of two-mode linear interface}\label{classification_begin}

We consider two modes, denoted by the annihilation operators $\hat{a}_1$ and $\hat{a}_2$, interacting through a linear interface. By introducing the Hermitian quadrature operators via $\hat{a}\equiv (\hat{q}+i\hat{p})/\sqrt{2}$, the transformation induced by the interface is represented as
\begin{eqnarray}\label{block_matrix_start}
\begin{pmatrix}
\hat{q}_i^{\text{out}}\\
\hat{p}_i^{\text{out}}
\end{pmatrix} = \sum_{j=1}^{2} {\bf T}^{ij}
\begin{pmatrix}
\hat{q}_j^{\text{in}}\\
\hat{p}_j^{\text{in}}
\end{pmatrix}.
\end{eqnarray}
Here ${\bf T}^{ij}$ with $i,j \in \{1,2\}$ is the $2 \times 2$ sub-matrix of the $4\times 4$ real symplectic matrix ${\bf T}$; describes the reflection or transmission of quadratures via the interface when $i=j$ or $i\neq j$ respectively \cite{rmp2012}. $\hat{q}^{\text{in(out)}}$ and $\hat{p}^{\text{in(out)}}$ denote the quadrature operators before (after) the interface. We note that Eq.~\eqref{block_matrix_start} is a general description of linear interfaces, i.e. it covers both the ones induced by scattering processes and coherent interaction. For scattering-type, e.g. interaction of travel photons via an optical beam splitter, $\hat{a}^{\text{in}}$ and $\hat{a}^{\text{out}}$ represent respectively the input and output propagating mode operators; for coherent interaction, e.g. coherent mode coupling between superconducting cavities \cite{superconducting_cavities2019}, $\hat{a}^{\text{in}}$ and $\hat{a}^{\text{out}}$ are the initial and final time mode operators respectively, i.e., $\hat{a}^{\text{in}}\equiv \hat{a}(0)$ and $\hat{a}^{\text{out}}\equiv \hat{\mathcal{U}}^\dag\hat{a}(0)\hat{\mathcal{U}}$ for some evolution operator $\hat{\mathcal{U}}$. 

Any interface can be converted by suitable single-mode operations to one of the seven typical interfaces: Identity, QNDI, TMS, BS, sTMS, sQNDI and SWAP \cite{lau2019}, i.e.
\begin{eqnarray}\label{standardform}
{\bf L}^{\text{out}}_1 {\bf L}^{\text{out}}_2  {\bf T} {\bf L}^{\text{in}}_1 {\bf L}^{\text{in}}_2= {\bf \bar{U}}.
\end{eqnarray} 
Here ${\bf L}^{\text{in(out)}}_i$ is a $4\times 4$ matrix denoting the single-mode Gaussian transformation on the $i$-th mode and ${\bf \bar{U}}$ denotes the standard form of the interface that both reflection and transmission matrices are diagonal and the non-zero diagonal entries have equal magnitude. This form is significant not only because the typical interfaces are usually expressed in the standard form in the literature, but it is also useful in interface engineering. The operator and matrix representations of the standard forms are listed in Table~\ref{main_table}. 

Generally, if single-mode operations are applied before and after the interface, the overall transformation matrix will be altered. Moreover, one would expect that an interface can generate inter-mode correlations that cannot be affected by single-mode operations. Indeed, Ref.~\cite{lau2019} identified that the ranks of the reflection and transmission matrices, $n_R \equiv \text{rank}({\bf T}^{22}) \in\{0,1,2\}$ and $n_T \equiv \text{rank}({\bf T}^{21}) \in\{0,1,2\}$, are invariant under single-mode operations. For the purpose of transduction, they show that each class of interface, as classified by the different combinations of ranks, has different utility in engineering a perfect transduction.

In additional to the ranks, another invariant, the determinant of the transmission matrix ${\bf T}^{21}$,
\begin{eqnarray}
\chi \equiv \text{det}({\bf T}^{21}) \in(-\infty,\infty),
\end{eqnarray}
is recognized in Ref.~\cite{lau2019}, although it is not involved in the engineering of transduction. 
For our current purpose of interface engineering, however, we realize that $\chi$ is a good classifier because all interfaces that share the same $\chi$ and ranks are inter-convertible. Explicitly, any two interfaces $A$ and $A'$ with the same $\chi$ and ranks have the same standard form according to Eq.~\eqref{standardform}, i.e. ${\bf L}^{\text{out}}_1 {\bf L}^{\text{out}}_2  {\bf T}_A {\bf L}^{\text{in}}_1 {\bf L}^{\text{in}}_2={\bf L}^{\text{out}'}_1 {\bf L}^{\text{out}'}_2 {\bf T}_{A'} {\bf L}^{\text{in}'}_1 {\bf L}^{\text{in}'}_2= {\bf \bar{U}}$. The two interfaces can thus be inter-converted by single mode operations, i.e. $({\bf L}^{\text{out}'}_1 {\bf L}^{\text{out}'}_2)^{-1}{\bf L}^{\text{out}}_1 {\bf L}^{\text{out}}_2  {\bf T}_A {\bf L}^{\text{in}}_1 {\bf L}^{\text{in}}_2 ({\bf L}^{\text{in}'}_1 {\bf L}^{\text{in}'}_2)^{-1}= {\bf T}_{A'}$.

To observe the physical meaning of $\chi$, we consider that for a BS with angle $\theta \in (-\pi/2,\pi/2)$,
\begin{eqnarray}\label{chi_BS}
\chi_{\text{BS}} = \sin^2\theta.
\end{eqnarray}
In this case, $\chi_{\text{BS}}$ is the magnitude square of the transmittance of the BS. Moreover, we recognize that no information is transmitted through Identity interface, of which $\chi_{\text{I}}=0$. While all information is transmitted through SWAP, this interface has a $\chi_{\text{SWAP}}=1$. From these examples, we can observe that $\chi$ can characterize the strength of the transmission through an interface, and hence we will call $\chi$ the transmission strength.

We now discuss the meaning of $\chi$ of the remaining interfaces. For sTMS and TMS, their transmission strengths are given by
\begin{eqnarray}
\chi_{\text{sTMS}} &=& \cosh^2 r\\
\chi_{\text{TMS}} &=&-\sinh^2 r,
\end{eqnarray}
where $r \in (-\infty,\infty)$ is the TMS strength. sTMS is equivalent to cascading two operations: amplifying by TMS and transmitting via SWAP. $\chi_{\text{sTMS}}$ is always larger than $1$ and implies that the transmitted information via sTMS is amplified. For TMS, the transmitted mode is associated with noise \cite{caves1982} and the negative $\chi_{\text{TMS}}$ is the signature of this fundamental incorporated noise. Both QNDI ($n_T=1$) and Identity ($n_T=0$) have the zero transmission strength, $\chi_{\text{QNDI}}=\chi_{\text{I}}=0$, it is because no quantum information is transmitted through either of them. Identity interface obviously transfers no information. Even though QNDI ($n_T=1$) transmits the information of one quadrature, its quantum capacity vanishes because this transmission can be equivalently implemented by homodyne detection and transmitting the classical measurement outcome.
sQNDI ($n_R=1$), similar to SWAP ($n_R=0$), has infinite quantum
capacity \cite{Jiang2022} and completely transmits quantum information without adding noise, so it has $\chi_{\text{sQNDI}}=\chi_{\text{SWAP}}=1$. 

Interestingly, even though both QNDI and sQNDI involve a non-zero QND strength $\eta \in (-\infty, \infty)$, this parameter cannot be used as a classifier since it is not invariant under single-mode transformation. More explicitly, it is easy to show that
\begin{eqnarray}
{\bf S}_1(\gamma^{-1}) {\bf \bar{U}}^Q(\eta) {\bf S}_1(\gamma)&=& {\bf \bar{U}}^Q(\gamma\eta)\label{QND_squeezing}\\
{\bf S}_1(\gamma^{-1}) {\bf \bar{U}}^{SQ}(\eta) {\bf S}_2(\gamma)&=& {\bf \bar{U}}^{SQ}(\gamma\eta),\label{sQND_squeezing}
\end{eqnarray}
where ${\bf S}_i(\gamma)$ denotes the single-mode squeezing with strength $\gamma \in (-\infty, \infty)$ of mode $i$ \footnote{Its expression can be found in Appendix A}, ${\bf \bar{U}}^Q(\eta)$ and ${\bf \bar{U}}^{SQ}(\eta)$ denote respectively the standard form of QNDI and sQNDI with QND strength $\eta$. Eqs.~(\ref{QND_squeezing})-(\ref{sQND_squeezing}) show that the QND strength can always be manipulated by applying single-mode squeezing before and afterward the interface.

\begin{widetext}
\begin{table*} 
    \centering
     {\renewcommand{\arraystretch}{1.2}    
    \begin{tabular}{| >{\centering}m{1.5cm} |c|c|c|c|c|c|}
    \hline
    \multicolumn{3}{|c}{\multirow{2}{*}{{\bf Characterization}}} & \multicolumn{2}{|c|}{\multirow{2}{*}{{\bf Standard Form}}} & \multicolumn{2}{c|}{{\bf no. of interfaces}}\\
    \multicolumn{3}{|c}{} & \multicolumn{2}{|c|}{} & \multicolumn{2}{c|}{{\bf for protocol}}\\\hline
      \multirow{2}{*}{{\bf Class}} & {\bf Transmission } &  {\bf Characteristic }& \multirow{2}{*}{{\bf Operator}} & \multirow{2}{*}{{\bf Matrix}}  & \multirow{2}{*}{{\bf General}} & \multirow{2}{*}{{\bf Restricted }}\\ 
       & {\bf Strength} &  {\bf Parameters}& & &  & \\
      \hline
    TMS &  $\chi<0$ &  $\Lambda$ & $ \exp{\left(-i r(\hat{q}_1\hat{p}_2+\hat{p}_1\hat{q}_2)\right)}$ 
       & $\begin{pmatrix}
\cosh r{\bf I} & \sinh r {\bf Z} \\
\sinh r {\bf Z}  & \cosh r{\bf I}  \\
\end{pmatrix}$  &2 &4\\ 
      Identity ($n_T=0$) & $\chi=0$ &  $\Lambda$  & $\hat{I}$ 
       &  $\begin{pmatrix}
{\bf I} & 0 \\
0  & {\bf I}  \\
\end{pmatrix}$ &3 & 5\\ 
        QNDI ($n_T=1$)& $\chi=0$ &  $\Lambda \quad \& \quad \kappa$ & $\exp{\left(-i\eta \hat{q}_1\hat{p}_2\right)}$
        &  $\begin{pmatrix}
{\bf I} & \eta\frac{{\bf I - Z}}{2} \\
\eta \frac{{\bf I+Z}}{2}  & {\bf I}  \\
\end{pmatrix}$  &2 &4\\
       BS & $0<\chi<1$ & $\Lambda$ & $\exp{\left(i\theta(\hat{q}_1\hat{p}_2 - \hat{p}_1\hat{q}_2)\right)}$
        & $\begin{pmatrix}
\cos \theta {\bf I} & -\sin \theta {\bf I} \\
\sin \theta {\bf I}  & \cos \theta {\bf I}  \\
\end{pmatrix}$  &2 &4\\ 
      sQNDI ($n_R=1$)& $\chi=1$ &  $\Lambda$ & $\hat{\mathbb{S}}\exp{\left(-i\eta \hat{q}_1\hat{p}_2\right)}$
       & $\begin{pmatrix}
\eta \frac{{\bf I - Z}}{2} & {\bf I} \\
{\bf I}  & \eta \frac{{\bf I + Z}}{2}  \\
\end{pmatrix}$  &2 &4\\ 
      SWAP ($n_R=0$)& $\chi=1$ &   / & $\hat{\mathbb{S}} \equiv \exp{\left(i\frac{\pi}{2}(\hat{a}^\dag_1-\hat{a}^\dag_2)(\hat{a}_1-\hat{a}_2)\right)}$ 
        &  $\begin{pmatrix}
0 & {\bf I} \\
{\bf I}  & 0  \\
\end{pmatrix}$  &3 &3\\ 
       sTMS & $\chi>1$ &  $\Lambda$ & $\hat{\mathbb{S}}\exp{\left(-i r(\hat{q}_1\hat{q}_2+\hat{p}_1\hat{p_2})\right)}$ 
        & $\begin{pmatrix}
\sinh r {\bf Z} & \cosh r{\bf I} \\
\cosh r{\bf I}  & \sinh r {\bf Z}  \\
\end{pmatrix}$  &2 &4\\ 
 \hline
    \end{tabular}}
        \caption{ 
        Characterization of interfaces and requirements for interface engineering. The first and second columns respectively show the equivalent well-known operation for each class and the corresponding transmission strength, $\chi \equiv \text{det}({\bf T_{21}})$. The third column shows the additional characteristic parameters when squeezing is restricted to only one mode.
        Forth and fifth columns list the operator and matrix representations of the interfaces in the standard form. Here $r$ is the TMS strength, $\theta$ is the BS angle, $\eta$ is the QND strength, ${\bf I}$ and ${\bf Z}$ are the $2\times 2$ identity and Pauli-Z matrices respectively. The last two columns are the number of fixed interfaces required for engineering a interface in this class under the general and squeezing-restricted situations.
   }\label{main_table}
\end{table*}
\end{widetext}

\section{Interface engineering with arbitrary single-mode operations}\label{sec_setup}

Each quantum information process requires purpose-specific interfaces. According to our characterization in Sec.~\ref{classification_begin}, implementing them is equivalent to implement the target interfaces with the specific transmission strength $\chi_{\text{tgt}}$ and ranks $n_{R}^{\text{tgt}}$ and $n_{T}^{\text{tgt}}$. However, physical platforms are usually subjected to practical limitations, so the available interfaces may be restricted. 
Inspired by previous works \cite{lau2019, Jiang2021}, we develop a systematic scheme that can engineer an interface with arbitrary transmission strength and ranks by cascading multiple component interfaces, i.e. platform-available interfaces. In this section, we consider the situation that any single-mode rotation and squeezing can be implemented on both modes. Our scheme is designed to account for the most general restrictions of the available interface, as we assume that both the ranks and transmission strength of every component interface are fixed but known.

Before discussing the details, it is worth clarifying the differences between our scheme and other interface engineering techniques. First, both the strength and type of an interface can be modified respectively by Hamiltonian amplification \cite{Ge2020hamiltonian} and Lloyd-Braunstein protocol \cite{SethLloyd1998}. However, the required number of component interfaces in these schemes grows with the accuracy and strength of the desired interface. On the contrary, our scheme requires at most three component interfaces to exactly engineer an interface with any desired strength and type. Second, by Bloch-Messiah decomposition \cite{braunstein2005squeezing}, any two-mode interface can be decomposed into two BS and a round of single-mode squeezing in between. However, engineering interfaces by this method requires implementing BS with tunable angles, which is in stark contrast to our scheme that can utilize any non-trivial component interfaces with fixed strength.

\subsection{Two-interface setup} \label{combination_possibility}
\begin{figure}
    \centering
    \includegraphics[scale=0.4]{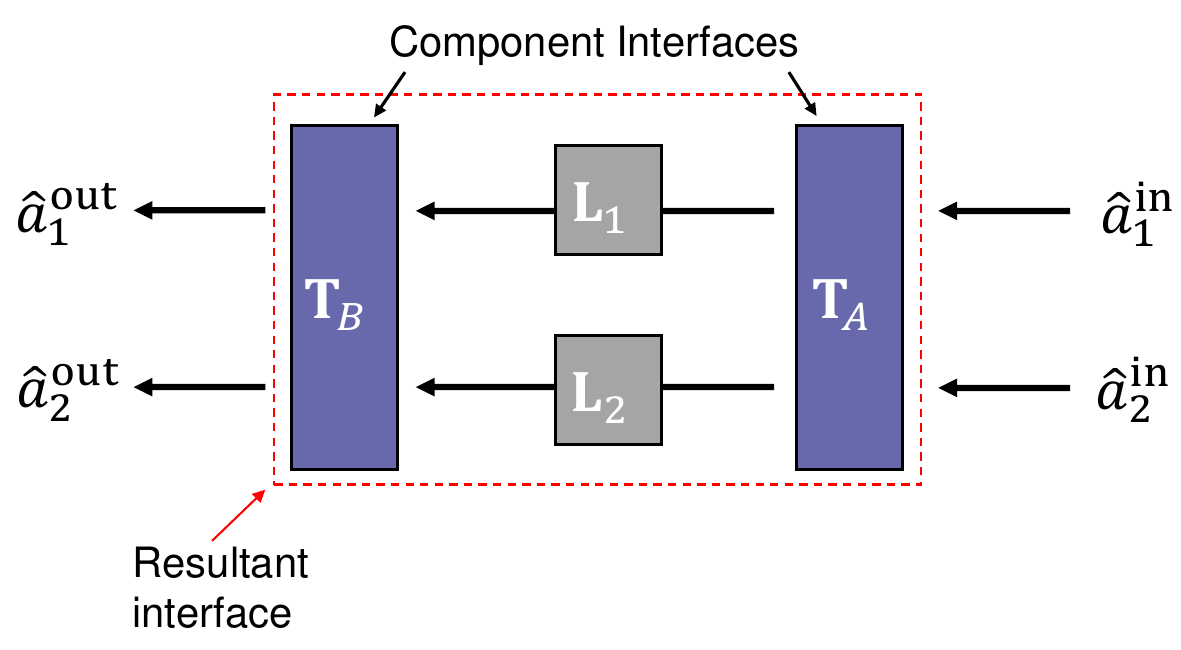}
    \caption{Two-interface setup for engineering a non-trivial interface. It consists of two two-mode component interfaces ${\bf T}_{A(B)}$ (blue) interspersed with the controllable $i$-th mode operations ${\bf L}_i$ (gray). The right-to-left black arrows represent the evolution of the bosonic mode operators. }
    \label{basical_setting}
\end{figure}

The basic setting consists of two cascading component interfaces as illustrated in Fig.~\ref{basical_setting}. This setup is implemented in the following sequence: 1) two modes couple via the first component interface $A$; 2) they are transformed by controllable single-mode operations, ${\bf L}_1$ and ${\bf L}_2$; 3) they couple again via the second component interface $B$. After the sequence, the two modes are effectively transformed under the resultant interface $AB$.

For simplicity, we assume that the component interfaces $A$ and $B$ have already been converted to the standard form by appropriate single-mode operations. Then, the resultant transformation matrix ${\bf T}_{AB}$ is given by, 
\begin{eqnarray}\label{general_setting}
{\bf T}_{AB} = {\bf \bar{U}}_B {\bf L}_1 {\bf L}_2 {\bf \bar{U}}_A.
\end{eqnarray}
Our aim is to find the suitable single-mode operations ${\bf L}_{1(2)}$, for constructing a resultant interface ${\bf T}_{AB}$ with the desired rank and transmission strength $\chi_{AB}$. 

In principle, there are 49 combinations of component interfaces because $A$ and $B$ can belong to any of the 7 classes in Table~\ref{main_table}. Fortunately, as explained below, only 6 combinations are independent and need to be considered individually. 

First, for engineering a non-trivial interface, it is legitimate to assume component interfaces $A$ and $B$ do not belong to Identity and SWAP, because these two classes do not generate any non-trivial correlation between the involving modes. 

Second, any combination involving sTMS or sQNDI components, which can be treated respectively as a TMS or QNDI followed by a SWAP, needs not be considered individually. We note that applying a SWAP after an interface with strength $\chi'$ will result in an overall interface with strength $1-\chi'$. It is because SWAP exchanges the transmitted and reflected quadratures, and hence their strengths, and the sum of transmission and reflection strengths always equal to unity according to the canonical commutation relations, i.e. $\det({\bf T}^{22})+\det({\bf T}^{21})=1$ \cite{lau2019}.
If the combination of $A$ and $B$ can engineer a resultant strength $\chi_{AB}$, then by using the same single-mode operations, an interface with strength $\chi_{AB'}=1-\chi_{AB}$ can be engineered by combining $A$ with a component $B'$ with strength $\chi_B'=1-\chi_B$, where $\chi_B$ is the transmission strength of $B$. 
As a result, the combination involving sTMS or sQNDI components can be deduced from the combinations of TMS or QNDI respectively.

Third, the order of interfaces does not affect the transmission strength. It can be understood by considering the inverse of Eq.~\eqref{general_setting}, i.e.
\begin{eqnarray}\label{order_relation}
({\bf \bar{U}}_A)^{-1} ({\bf L}_1 {\bf L}_2)^{-1} ({\bf \bar{U}}_B)^{-1} = ({\bf T}_{AB})^{-1},
\end{eqnarray}
and recognizing that any interface has the same transmission strength as its inverse. For any non-trivial interface, its inverse can be constructed by applying single-mode operations before and after itself, i.e.
\begin{eqnarray}
&&\left({\bf \bar{U}}^{\chi}\right)^{-1} = \begin{cases}
{\bf R}_1(\pi) {\bf \bar{U}}^{\chi} {\bf R}_1(\pi) & \text{for}\quad \chi<1\\
{\bf R}_{1(2)}(\pi) {\bf \bar{U}}^{\chi} {\bf R}_{2(1)}(\pi) & \text{for}\quad \chi>1
\end{cases},\label{U_chi}\\
&&\left({\bf \bar{U}}^{Q}(\eta)\right)^{-1} = {\bf R}_1(\pi) {\bf \bar{U}}^{Q}(\eta) {\bf R}_1(\pi),\label{U_Q}\\
&&\left({\bf \bar{U}}^{SQ}(\eta)\right)^{-1} = {\bf F}^{-1}_1 {\bf F}_2{\bf \bar{U}}^{SQ}(\eta){\bf F}_1 {\bf F}_2^{-1},\label{U_SQ}
\end{eqnarray}
where ${\bf R}_i$ and ${\bf F}_i \equiv {\bf R}_i(\pi/2)$ denote respectively the rotation and Fourier gate for the $i$-th mode \footnotemark[\value{footnote}]. The standard form of BS, TMS and sTMS interfaces are denoted as ${\bf \bar{U}}^\chi$. If there exists a protocol that engineers a desired interface $AB$ by first applying $A$ then $B$ (denoted by combination $A+B$ hereafter), one can substitute Eqs.~(\ref{U_chi})-(\ref{U_SQ}) into Eq.~\eqref{order_relation} to obtain a protocol to generate $AB$ by first applying $B$ then $A$.

As a consequence, we need to consider only six combinations: BS+BS, TMS+TMS, QND+QND, BS+TMS, BS+QND and TMS+QND.

\subsection{Interference mechanism}

The principle of the cascading interface setup is to use the controllable single-mode operations ${\bf L}_{1(2)}$ to interfere and hence manipulate the transmission. As a pedagogical example, we consider a simple setting with only one controllable squeezing in between, i.e. ${\bf L}_1= {\bf S}_1(\gamma)$ and ${\bf L}_2={\bf I}$.
Furthermore, we assume 
the component interfaces $A$ and $B$ are respectively BS and TMS in the standard form. Since both interfaces and single-mode operations are quadrature-diagonal, the overall transmission matrix is given by
\begin{eqnarray}
{\bf T}_{AB}^{21} =\begin{pmatrix}
T^{21,q}_{AB} & 0\\
0 & T^{21,p}_{AB}
\end{pmatrix},
\end{eqnarray}
where 
\begin{eqnarray}
T^{21,q}_{AB} &=&\gamma \sqrt{\chi_B(\chi_A-1)} - \sqrt{(1-\chi_B)\chi_A},\label{Tq}\\
T^{21,p}_{AB} &=&-\gamma^{-1} \sqrt{\chi_B(\chi_A-1)}-\sqrt{(1-\chi_B)\chi_A}.\label{Tp}
\end{eqnarray}

\begin{figure}
    \centering
    \includegraphics[scale=0.85]{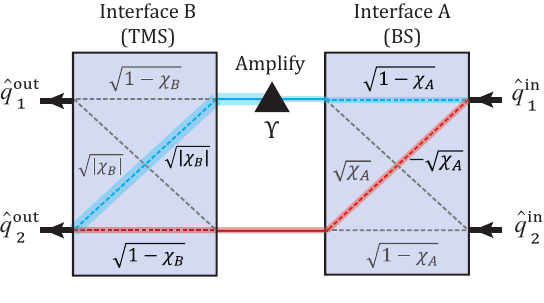}
    \caption{Interference of $q$-quadrature transmission for the cascading interfaces setup. There are two paths for transmitting $\hat{q}^{\text{in}}_1$ to $\hat{q}^{\text{out}}_2$. Due to the squeezing of mode $1$ (triangle), the blue path is amplified and the red one is unchanged. The overall transmission amplitude is determined by the interference of the two paths and thus controllable by the squeezing strength $\gamma$.}
    \label{interference_1}
\end{figure}

As illustrated in Fig.~\ref{interference_1}, the $q$-transmission amplitude of $\hat{q}_1$ to $\hat{q}_2$ is the sum of the amplitudes of two paths: the upper path (blue) reflects in $A$, passes through the squeezer and transmits in $B$; and the lower path (red) transmits in $A$ and reflects in $B$. The squeezer amplifies the amplitude of the first path, i.e. the first term in Eq.~(\ref{Tq}), while the amplitude of the second path is not modified, i.e. the second term in Eq.~(\ref{Tq}). A similar process happens for the $p$-transmission in Eq.~\eqref{Tp}, except that the first path is deamplified by the squeezer. It is clear that adjusting the squeezing strength $\gamma$ can tune the resultant transmission strength $\chi_{AB}=T_{AB}^{21,q}T_{AB}^{21,p}$. $\chi_{AB}$ can be arbitrarily small near destructive interference, i.e. a $\gamma$ is chosen that $T_{AB}^{21,q}$ or $T_{AB}^{21,p}$ is close to zero, while it can be arbitrarily large by exerting a large $\gamma$ such that $T_{AB}^{21,q}$ is large. As a result, the whole spectrum of $\chi_{AB}$ can be covered by manipulating $\gamma$. 

\subsection{General Two-interface Protocols}\label{3_configuration}

Next, we present the complete protocols for all six combinations of component interfaces. We will consider the setting that consists of mode-1 squeezing and rotations on both modes, i.e. ${\bf L}_1 = {\bf R}_1(\phi_1){\bf S}_1(\gamma)$ and ${\bf L}_2 = {\bf R}_2(\phi_2)$, where $\phi_1,\phi_2 \in (0,2\pi)$ are the rotation angles. A straightforward calculation shows that the resultant transmission strength is given by
\begin{eqnarray}\label{chi_BA}
\chi_{AB} &\equiv& \text{det}\left[ [{\bf \bar{U}}_B {\bf R}_1(\phi_1){\bf S}_1(\gamma) {\bf R}_2(\phi_2) {\bf \bar{U}}_A]^{21} \right]  \nonumber\\
&=& X_f + Z,
\end{eqnarray}
where $X_f \equiv \chi_{B} + \chi_{A} - 2  \chi_{B} \chi_{A}$ is determined only by the transmission strength of the component interfaces, and $Z$ accounts for the interference that is controllable by the in-between single-mode operations. Our strategy is to tune $Z$ to adjust $\chi_{AB}$ to the desired value, $\chi_{\text{tgt}}$. $Z$ takes different forms for different combinations. We will discuss each combination as follows, and the result is summarized in Table~\ref{all_combination}.

 \begin{table}
    \centering
\begin{tabular}{|c|c|c|c|c|c|}
    \hline
       Config. & {\footnotesize Identity} & {\footnotesize QNDI} & {\footnotesize BS/TMS/sTMS} & {\footnotesize sQNDI} & {\footnotesize SWAP}\\\hline 
       {\scriptsize BS+BS} & $\times^*$ & \checkmark & \checkmark & \checkmark& $\times^\dag$\\ 
       {\scriptsize TMS+TMS }& $\times^*$ & \checkmark&  \checkmark & \checkmark& $\times$\\
        {\scriptsize QNDI+QNDI }& \checkmark &\checkmark &\checkmark & \checkmark&  $\times$\\
        {\scriptsize BS+TMS }& $\times$ & \checkmark&\checkmark& \checkmark& $\times$\\
       {\scriptsize BS+QNDI }&$\times$ & \checkmark& \checkmark&\checkmark & $\times$\\
        {\scriptsize TMS+QNDI }& $\times$ & \checkmark& \checkmark& \checkmark& $\times$\\\hline
    \end{tabular}
    \caption{Possibility of engineering each class of interface with the six combinations of two component interfaces. For the resultant interface being BS/TMS/sTMS, it is always possible to engineer $\chi$ to any desired value.  $^*$ It is generally not possible except the special case that the two component interfaces have the same strength, $\chi_A = \chi_B$. $^\dag$ It is generally not possible except the special case that the two component interfaces have complementary strengths, $\chi_A = 1- \chi_B$.}
    \label{all_combination}
\end{table}

\subsubsection{BS+TMS, BS+QNDI or TMS+QNDI}\label{protocol_diff}

When two interfaces of different classes are combined, we turn off all the rotations, i.e. $\phi_1=\phi_2=0$, and the value of $Z$ is solely determined by the squeezing strength $\gamma$. For BS+TMS, $Z$ is given by
\begin{eqnarray}
Z_\textrm{BS+TMS} &=& \sqrt{\chi_{B}(1-\chi_{B})\chi_{A}(\chi_{A}-1)}\frac{\gamma^2-1}{\gamma}, \label{Z_BT}
\end{eqnarray}
which can be tuned to arbitrary values. For BS+QNDI and TMS+QNDI, their $Z$ are expressed as
\begin{eqnarray}
Z_\textrm{BS+QNDI} &=& -\sqrt{\chi_{B}(1-\chi_{B})}\eta\gamma^{-1}, \label{Z_BQ}\\
Z_\textrm{TMS+QNDI} &=& -\sqrt{\chi_{B}(\chi_{B}-1)}\eta\gamma^{-1} \label{Z_TQ},\end{eqnarray}
which can be arbitrary non-zero value \footnote{$Z=0$ is useless for these two combinations of component interfaces. It is because $\chi_{BA}=\chi_B + Z$ for these cases and engineering $Z=0$ just gives us the interface $B$ which we already have.}. Under these settings, the resultant interface can be any non-trivial interface, i.e. BS, TMS, sTMS, QNDI or sQNDI, but not Identity and SWAP. We will discuss more about the latter two in Sec.~\ref{engineer_Iden_SWAP}.

\subsubsection{BS+BS or TMS+TMS}\label{protocolBBTT}

For the combination BS+BS or TMS+TMS with $\chi_A \neq \chi_B, 1-\chi_B$, we can set $\phi_2=0$, and obtain $Z$ as
\begin{eqnarray}
Z_\textrm{BS+BS} &=& \frac{Z_0}{2}\frac{\gamma^2+1}{\gamma}\cos\phi_1, \label{Z_BB}\\
Z_\textrm{TMS+TMS} &=& -\frac{Z_0}{2}\frac{\gamma^2+1}{\gamma}\cos\phi_1, \label{Z_TT}
\end{eqnarray}
where $Z_0\equiv 2\sqrt{\chi_{B}(1-\chi_{B})\chi_{A}(1-\chi_{A})}$.

Although $|(\gamma^2+1)/\gamma|$ is lower bounded by 2, $Z$ covers all real values thanks to the rotation $\phi_1$. More explicitly,
\begin{enumerate}
\item When $\chi_{\text{tgt}}$ falls between 
$X_f -Z_0 < \chi_{\text{tgt}} < X_{f} + Z_0$, we turn off the squeezing, i.e. $\gamma=1$, and adjust the value of $Z$ by $\phi_1$.

\item Otherwise, i.e. $ \chi_{\text{tgt}}<X_{f} -Z_0$ or $\chi_{\text{tgt}}>X_{f} + Z_0$, we set $\phi_1=0$ and manipulate the value of $Z$ by $\gamma$. 
\end{enumerate}
Under these settings, we can engineer any non-trivial interface.

We now discuss two special cases. First, for $\chi_A=\chi_B$, we will obtain Identity instead of QNDI when $\chi_{AB}=0$. To engineer QNDIs, a rotation is needed to mix the $q$- and $p$-quadratures to avoid the complete destructive interference of both quadratures in the transmission. We show in Appendix B that QNDIs can be constructed by setting $\phi_1\neq 0$, $\phi_2=0$ and $\gamma = \tan\phi_1 - \sec\phi_1$.

Another special case is $\chi_A=1-\chi_B$, which may happen in BS+BS combination. Tuning $\chi_{AB}=1$ by the above protocol will result in a SWAP but not sQNDI. To construct sQNDIs, we need a modified setting to prevent reflection from completely destructively interfering in both quadratures. We show in Appendix B that this can be achieved by setting $\phi_1 \neq 0$, $\phi_2=0$ and $\gamma=-\tan\phi_1-\sec\phi_1$. 

\subsubsection{QNDI+QNDI}\label{protocol_QND}

If one of the available component interfaces is a QNDI, it can be used to engineer a QNDI with arbitrary QND strength by applying single-mode squeezing, i.e. Eq.~\eqref{QND_squeezing}.
For other non-trivial interfaces, we can pick $\phi_1=\phi_2=\pi/2$ and obtain
\begin{eqnarray}
Z_{\text{QNDI}+\text{QNDI}} &=& - \eta_A\eta_B\gamma^{-1},\label{Z_QQ}
\end{eqnarray}
which can be adjusted by $\gamma$ to cover all non-zero values \footnote{The resultant transmission strength is equal to exactly $Z$ for this case, so it is not necessary to engineer $Z=0$ which corresponds to the QNDI}.  
We are also able to engineer an Identity by changing the setting to $\phi_1 =\phi_2 = 0$ and $\gamma=-\eta_A/\eta_B$, then the two component QNDIs will be mutually cancelled. 

\subsection{Three-interface Protocols for Identity and SWAP}\label{engineer_Iden_SWAP}

In view of the success of two-interface protocol in engineering non-trivial interface, one might be tempted to use such protocol to also engineer Identity and SWAP. Engineering these interfaces requires eliminating the amplitudes of both $q$- and $p$-quadratures in either the transmission (for Identity) or reflection (for SWAP). However, in the two-interface configuration, there is only one round of squeezing between the two component interfaces. Intuitively, it seems impossible to simultaneously destructively interfere both $q$- and $p$- quadratures as the squeezing of one quadrature will inevitably anti-squeeze the other. 

To verify this intuition, we exploit all available single-mode controls in Appendix B, and discover that Identity and SWAP cannot be engineered by the two-interface protocol unless the two interfaces follow stringent relations. For engineering Identity, the component interfaces must have the same strength, $\chi_A = \chi_B $. In this situation, interface $A$ can always be converted to the inverse of $B$ by appropriate single-mode operations (see Eqs.~(\ref{U_chi})-(\ref{U_SQ})), then the two component interfaces mutually cancel and the resultant interface becomes an Identity.

To engineer SWAP, the strengths of the component interfaces must be complemented, i.e. $\chi_A = 1-\chi_B$. This condition implies that interface A can be converted to the swapped-inverse of interface B, i.e. ${\bf T}_A={\bf T}^{-1}_B {\bf \bar{U}}^S$, where ${\bf \bar{U}}^S$ is the standard form of SWAP. By eliminating ${\bf T}_B$, we obtain a SWAP.

\begin{figure}
    \centering
    \includegraphics[scale=0.4]{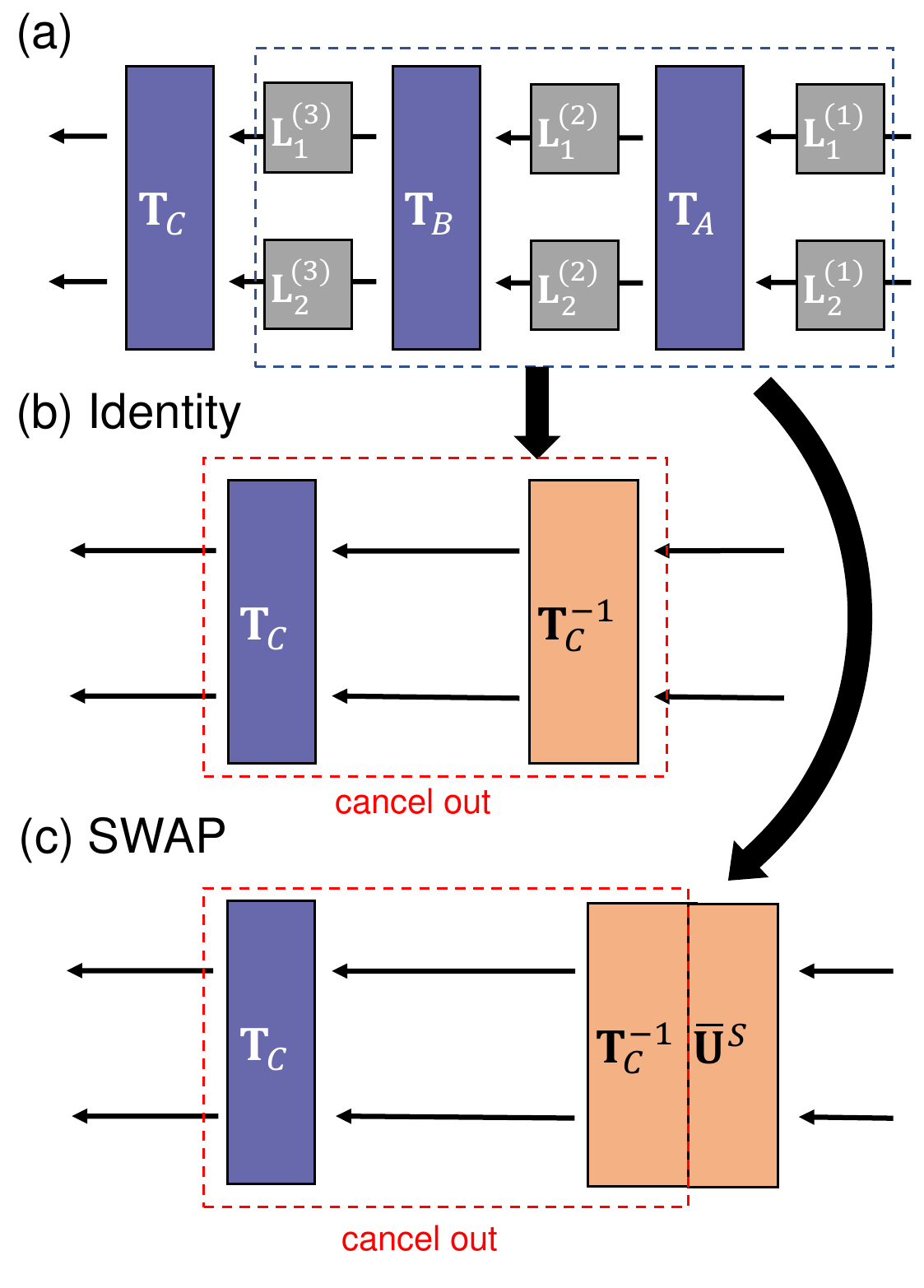}
    \caption{Three-interface protocol for engineering Identity and SWAP. (a) The scheme involves cascading three component interfaces $A$, $B$, and $C$ and suitable single-mode controls. (b) For engineering Identity, $A$ and $B$ are utilized to engineer an intermediate interface $AB$ to be the inverse of $C$. (c) For constructing SWAP, $AB$ is engineered to be the swapped-inverse of $C$.} \label{three_interfaces}
\end{figure}

In general, the available component interfaces may not satisfy the above stringent conditions. To engineer Identity and SWAP, our strategy is to cascade more interfaces and single-mode controls, such that additional rounds of interference can be implemented to destructively interfere both quadratures. Inspired by the above understandings, we develop systematic protocols that require the minimum rounds of component interfaces, i.e. using three interfaces, $A$, $B$ and $C$, as illustrated in Fig.~\ref{three_interfaces}(a). 

For engineering Identity, our strategy is to apply the two-interface protocol in Sec.~\ref{3_configuration} to combine two component interfaces, $A$ and $B$, to construct an intermediate interface $AB$ that is the inverse of $C$ (Fig.~\ref{three_interfaces}(b)). Similarly, to engineer SWAP, $AB$ can be engineered to be the swapped-inverse of $C$, i.e. ${\bf T}_{AB}={\bf T}_{C}^{-1}{\bf \bar{U}}^{S}$, so that the resultant interface $ABC$ becomes a SWAP (Fig.~\ref{three_interfaces}(c)). 

To summarize this section, any non-trivial interface, BS, TMS, sTMS, QNDI and sQNDI, can be engineered by cascading two component interfaces with suitable single-mode controls. Moreover, engineering Identity and SWAP is generally impossible with only two arbitrary interfaces, so we have introduced a three-interface protocol. Comparing with the existing schemes that require six arbitrary component interfaces to engineer a SWAP \cite{lau2019}, and four identical component interfaces for engineering an arbitrary two-mode interface \cite{Jiang2021}, our protocols are optimal because the number of component interface involved is the minimum possible. 

\section{Characterizing Squeezing Restricted interface}\label{sit_with_squeezing_restriction}

So far we have considered the situation that all single-mode operations are available on both modes. However, this situation is not always relevant in practice because actively squeezing an unknown state is known to be challenging in many platforms, such as optics and spin ensembles. If one involved platform in a hybrid quantum system suffers from such squeezing restriction, one would expect that the interfaces with different squeezing on the restricted mode cannot be inter-converted by the available single-mode controls. 

Apart from hybrid quantum systems, such squeezing restriction can also be found even when all modes have the same physical nature, but each mode plays a different role in the application.
For example, in collision-based trapped ion cooling \cite{rapid_cooling_ion2023}, which aims to cool data ions that encoded information in their internal states, squeezing can only be applied to the ancillary coolant ions that do not encode information. It is because squeezing ion motion would require coupling the motional and internal states \cite{motional_ion1996}, and thus will corrupt the encoded information.

Although the squeezing restriction is not uncommon, its implications on interface characterization and engineering are not well understood. In the relevant literature, Ref.~\cite{lau2019} studies this issue only for the engineering of a specific interface (i.e. SWAP). The interface engineering scheme in Ref.~\cite{Jiang2021}, Moreover, heavily relies on the phase-sensitive amplification of every mode and thus cannot be applied under this practical restriction. We will fill this missing gap in the following sections. In this section, we will first complete the interface characterization, and its implication in interface engineering will be discussed in Sec.~\ref{modified_protocol}.

\subsection{Diagonalizing mode-2 reflection matrix ${\bf T}^{22}$}

In the previous sections, we illustrate that each interface can be classified by its transmission strength and sub-matrix ranks under the assumption that arbitrary Gaussian single-mode operation can be applied. Under the squeezing restriction, however, these classifiers are no longer sufficient.  Here is an example: consider a linear interface that introduces only single-mode squeezing to the restricted mode, this interface falls into the class of Identity. Obviously, this interface cannot be converted to some other interface belongs to the same class, e.g. an operation that is identity on both modes, because it requires squeezing of the restricted mode. Thus a modified scheme is needed to classify the interfaces under squeezing restriction, and we will introduce it as follows.

Without loss of generality, we assume mode 2 is squeezing-restricted, the allowable operation in this setup will consist of any single-mode operation on mode 1 but only rotation on mode 2.
We first consider the mode-2 reflection matrix ${\bf T}^{22}$ since it is the part in the transformation matrix ${\bf T}$ that is mostly affected by this restriction, i.e. ${\bf T}^{22}$ can be altered by rotation only. 
According to the property of singular value decomposition (SVD) \cite{qu_decomp}, the singular values of a matrix are invariant under rotation operations. The SVD of ${\bf T}^{22}$ is given by 
\begin{eqnarray}\label{T22_to_Sigma}
{\bf T}^{22} = [({\bf R}^{\text{out}}_{2})^{-1}]^{22}{\bf \Sigma}[({\bf  R}^{\text{in}}_{2})^{-1}]^{22},
\end{eqnarray}
where ${\bf \Sigma}$ is a $2 \times 2$ diagonal matrix with the singular values. 
We thus realize that the singular values of ${\bf T}^{22}$ can be used to construct the invariant classifier under squeezing restriction. We will discuss its physical meaning for each class of interface.

\subsubsection{Interfaces with $\chi \neq 0,1$}\label{interface_not_0_1}

For interfaces with $\chi \neq 0,1$, the corresponding ${\bf \Sigma}$ can always be written as
\begin{eqnarray}\label{Sigma}
{\bf \Sigma}^\chi \equiv \begin{pmatrix}
\Lambda \sqrt{|1-\chi|} & 0 \\
0 & \pm\Lambda^{-1} \sqrt{|1-\chi|}
\end{pmatrix},
\end{eqnarray}
where the $+$ and $-$ signs are respectively for the interfaces with $\chi<1$ and $\chi>1$. The additional classifier $\Lambda$ characterizes the discrepancy between the two singular values. We identify $\Lambda$ as the irreducible squeezing strength since it will not be altered unless squeezing is applied on mode 2. We can denote $\Lambda$ as the squeezing of mode 2, i.e. $\Lambda \geq 1$, by setting ${\bf \Sigma}^\chi$ with the positive and larger upper diagonal element, since the squeezed and anti-squeezed quadratures of mode 2 can always be exchanged by applying Fourier gates before and after the interface.

In Appendix C, we show that the transformation matrix ${\bf T}$ of any interface in this class can always be transformed, by the allowable single-mode operations, into one of the following two simplest forms that consist of only two irreducible parts: a standard-form interface and a mode-2 squeezing. They are the pre-squeezing form, i.e. mode-2 squeezing is applied before a standard-form interface,
\begin{eqnarray}
{\bf L}^{\text{out}}_1{\bf R}^{\text{out}}_2{\bf T}{\bf L}_1^{\text{in}}{\bf R}^{\text{in}}_2 &=& {\bf \bar{U}^\chi }{\bf S}_2 (\Lambda)\label{M_R_chi},
\end{eqnarray}
and the post-squeezing form, i.e. mode-2 squeezing is applied after a standard-form interface,
\begin{eqnarray}
{\bf L}^{\bar{\text{out}}}_1{\bf R}^{\bar{\text{out}}}_2{\bf T}{\bf L}_1^{\bar{\text{in}}}{\bf R}^{\bar{\text{in}}}_2  &=& {\bf S}_2 (\Lambda) {\bf \bar{U}^\chi }.\label{M_L_chi}
\end{eqnarray}
They are useful for constructing the squeezing-restricted protocols in Sec.~\ref{modified_protocol}-\ref{remote_gate} and show clearly that each interface in this class is characterized by two invariant parameters: the transmission strength $\chi\neq 0,1$, and the irreducible squeezing $\Lambda \geq 1$. 

\subsubsection{sQNDIs ($\chi=1$, $n_R=1$)}

For interfaces with $\chi=1$ and $n_R = 1$, their ${\bf \Sigma}$ is of rank 1 and given by
\begin{eqnarray}\label{sigma_sq}
{\bf \Sigma}^{SQ} \equiv \begin{pmatrix}
\Lambda  & 0 \\
0 & 0
\end{pmatrix}.
\end{eqnarray}
Here the non-zero singular value $\Lambda$ can be recognized as the QND strength of the sQNDI. Different from $\Lambda \geq 1$ for interfaces with $\chi \neq 0,1$ (Sec.~\ref{interface_not_0_1}), $\Lambda$ for sQNDIs should denote both squeezing or anti-squeezing of mode 2, i.e. $\Lambda>0$. It is because only one quadrature is presented in the reflection through a sQNDI, applying Fourier gates cannot switch the squeezing to anti-squeezing of this quadrature nor vice versa. Moreover, in contrast to the situation that both modes can be squeezed, the QND strength is invariant under the squeezing restriction. This can be understood from Eq.~\eqref{sQND_squeezing} that altering the QND strength of a sQNDI requires squeezing both modes, but since it is forbidden in this situation, the QND strength becomes invariant.  We show in Appendix C that any interface in this class can be transformed into the pre- and post-squeezing forms,
\begin{eqnarray}
{\bf L}^{\text{out}}_1{\bf R}^{\text{out}}_2{\bf T}{\bf L}_1^{\text{in}}{\bf R}^{\text{in}}_2&=& {\bf \bar{U}}^{SQ} (1){\bf S}_2 (\Lambda),  \label{M_R_SQ}\\ 
{\bf L}^{\bar{\text{out}}}_1{\bf R}^{\bar{\text{out}}}_2{\bf T}{\bf L}_1^{\bar{\text{in}}}{\bf R}^{\bar{\text{in}}}_2   &=& {\bf S}_2 (\Lambda) {\bf \bar{U}}^{SQ}(1).\label{M_L_SQ}
\end{eqnarray}
Both forms consist of a sQNDI with a unity strength and a mode-2 squeezing with strength $\Lambda$. The latter can be recognized as an irreducible squeezing and it can be used to characterize a sQNDI.

\subsubsection{QNDIs ($\chi=0$, $n_T=1$)}

For interfaces with $\chi=0$ and $n_T=1$, their ${\bf T}^{22}$ can always be diagonalized by SVD as
\begin{eqnarray}
{\bf \Sigma}^Q \equiv \begin{pmatrix}
\Lambda & 0\\
0 & \Lambda^{-1}
\end{pmatrix}.
\end{eqnarray}
Same as Sec.~\ref{interface_not_0_1}$, \Lambda \geq 1$ can be recognized as irreducible squeezing and can be used to characterize QNDIs. Surprisingly, in addition to $\Lambda$, we discover another parameter from the transmission matrix ${\bf T}^{21}$ that is invariant under the allowable operations. We show in Appendix C that the transmission matrix can be decomposed as ${\bf T}^{21} = [({\bf R}_2^{\text{out}})^{-1}]^{22} {\bf N} [({\bf L}_1^{\text{in}})^{-1}]^{11}$ by QL decomposition \cite{qu_decomp}, where
 \begin{eqnarray}\label{N_eq}
{\bf N} \equiv \begin{pmatrix}
\eta  & 0 \\
\eta \kappa & 0
\end{pmatrix}=\begin{pmatrix}
1  & 0 \\
 \kappa & 1
\end{pmatrix}\begin{pmatrix}
\eta  & 0 \\
0 & 0
\end{pmatrix}
\end{eqnarray}
Here $\eta$ is the QND strength, which can be manipulated by squeezing mode 1. Moreover, the parameter $\kappa$ is invariant under the available single-mode operations. We call $\kappa$ the irreducible shearing \cite{rmp2012} since it can be recognized as an additional mode-2 shearing applied after the standard-form QNDI. 

Since a shearing operation can be decomposed into rotation and squeezing \cite{rmp2012}, in Appendix C we show that any QNDI can be transformed to the following pre- and post-squeezing forms \footnote{The pre- or post-squeezing form of QNDI has the non-diagonal mode-2 reflection matrix, so the required single-mode operations are generally different from that for diagonalizing ${\bf T}^{22}$ (Eq.~\eqref{T22_to_Sigma}) and transforming ${\bf T}^{21}$ into ${\bf N}$ (Eq.~\eqref{N_eq}).},
\begin{eqnarray}
{\bf L}^{\tilde{\text{out}}}_1{\bf R}^{\tilde{\text{out}}}_2{\bf T}{\bf L}_1^{\tilde{\text{in}}}{\bf R}^{\tilde{\text{in}}}_2&=&{\bf \bar{U}}^Q(\eta) {\bf R}_2(\phi_R){\bf S}_2(\Lambda),\label{M_R_Q}
\\
{\bf L}^{\bar{\text{out}}}_1{\bf R}^{\bar{\text{out}}}_2{\bf T}{\bf L}_1^{\bar{\text{in}}}{\bf R}^{\bar{\text{in}}}_2 &=&{\bf S}_2(\Lambda) {\bf R}_2(\phi_L){\bf \bar{U}}^Q(\eta)\label{M_L_Q}.
\end{eqnarray}
Here $\tan\phi_R \equiv \kappa$ and $\tan\phi_L \equiv -\kappa \Lambda^2$.
Overall, any QNDI is characterized by three parameters, $\chi=0$, $\Lambda$ and $\kappa$. 

\subsubsection{Identity ($\chi=0$, $n_T=0$)}

Any interface with $\chi=0$ and $n_T=0$ is equivalent to single-mode operations applying to both modes.  Under the squeezing restriction, any operation on the unrestricted mode $1$ can be removed by single-mode controls, while that on the squeezing-restricted mode $2$ can be removed up to a squeezing operation, i.e.
\begin{eqnarray}
{\bf L}^{\text{out}}_1{\bf R}^{\text{out}}_2{\bf T}{\bf L}_1^{\text{in}}{\bf R}^{\text{in}}_2 &=& {\bf S}_2 (\Lambda).\label{M_I}
\end{eqnarray}
In other words, ${\bf T}^{22}$ can be diagonalized according to Eq.~\eqref{T22_to_Sigma} with ${\bf \Sigma} = {\bf S}^{22}_2(\Lambda)$.
As such, any interface in this class is characterized by the irreducible squeezing strength $\Lambda$.

\subsubsection{SWAP ($\chi=1$, $n_R=0$)}

Any interface with $\chi=1$, $n_R=0$ can always be converted to a standard-form SWAP by the available single-mode operations, i.e.
\begin{eqnarray}\label{M_swap}
{\bf L}^{\text{out}}_1{\bf R}^{\text{out}}_2{\bf T}{\bf L}_1^{\text{in}}{\bf R}^{\text{in}}_2 &=& {\bf \bar{U}}^S .
\end{eqnarray}
There is no irreducible squeezing for this class since any squeezing applying on mode $2$ can be swapped to mode $1$, i.e. ${\bf \bar{U}}^S {\bf S}_2(\Lambda) ={\bf S}_1(\Lambda){\bf \bar{U}}^S $, and be cancelled by mode-$1$ squeezing controls.

We summarize all additional characteristic parameters due to the squeezing restriction for every class of interface in Table.~\ref{main_table}. We note that, in analogous to the unrestricted case, any two interfaces $A$ and $A'$ that share the same set of characteristic parameters are inter-convertible. It is because they can be transformed by the allowable single-mode operations to the same standard form in conjunction with the same irreducible local operations (Eqs.~(\ref{M_R_chi}), (\ref{M_L_chi}), (\ref{M_R_SQ}), (\ref{M_L_SQ}), (\ref{M_R_Q}), (\ref{M_L_Q}), (\ref{M_I}), (\ref{M_swap})), i.e. ${\bf L}^{\text{out}}_1 {\bf R}^{\text{out}}_2  {\bf T}_A {\bf L}^{\text{in}}_1 {\bf R}^{\text{in}}_2 = {\bf L}^{\text{out}'}_1 {\bf R}^{\text{out}'}_2 {\bf T}_{A'} {\bf L}^{\text{in}'}_1 {\bf R}^{\text{in}'}_2 $. Inter-conversion can then be done by inverting the single-mode operations.

\section{Squeezing-restricted Protocols}\label{modified_protocol}

Because each interface possesses additional invariant properties under the squeezing restriction, one might expect that the protocols in Sec.~\ref{sec_setup} are no longer general in engineering arbitrary interfaces in this situation.  Indeed, there are two main reasons that a modified interface engineering protocol is needed:
1) in addition to ranks and transmission strength, the protocol should also generate a target interface with the desired irreducible squeezing and shearing; 2) the irreducible properties of the component interfaces have to be considered. In the following, we will present the modified protocols that take these two issues into account. 
Intuitively, in order to cope with the additional restrictions, the number of the required component interfaces will increase; such requirement is summarized in Table~\ref{main_table}.

\subsection{Two-interface module}\label{two_interface_module}

\begin{figure}
    \centering
    \includegraphics[scale=0.45]{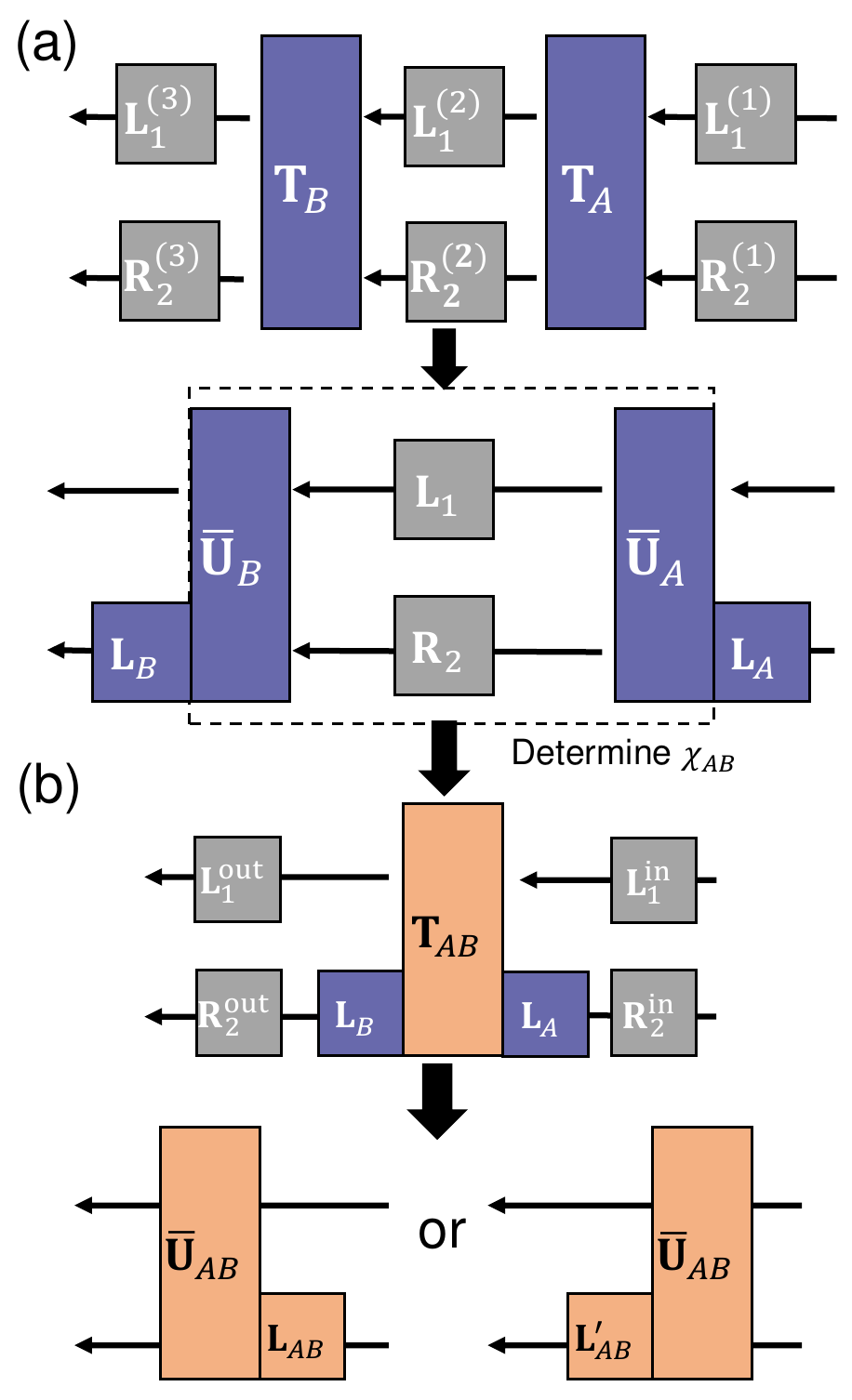}
    \caption{Two-interface module for engineering an intermediate interface under squeezing restriction. (a) Component interfaces $A$ and $B$ are first converted to respectively the pre- and post-squeezing forms, and then combined to form an intermediate interface $AB$ with strength $\chi_{AB}$ by using the protocol in Sec.~\ref{two_interface_module}. (b) With suitable single-mode operations, interface $AB$ can be transformed to the pre-squeezing (or post-squeezing) form for further processing.}
    \label{modified_two_interface_protocol}
\end{figure}

We start with introducing the two-interface module that is required in the protocols. Assuming a component interface $A$ is applied before $B$, the first step is to convert $A$ and $B$ into respectively the pre- and post-squeezing forms, as illustrated in Fig.~\ref{modified_two_interface_protocol}(a), such that the irreducible mode-2 operations are placed at the very beginning and very end of the whole sequence. By considering the standard-form parts of $A$ and $B$, we can then apply the two-interface protocols in Sec.~\ref{3_configuration} to engineer an intermediate interface $AB$ with the desired $\chi_{AB}$, which is not affected by the irreducible squeezing and shearing of $A$ and $B$. After that, we can convert $AB$ into either the pre- or post-squeezing form (Fig.~\ref{modified_two_interface_protocol}(b)), depending on the context of the protocol. 

\subsection{SWAP}\label{swap_engineering_restriction}

\begin{figure}
    \centering
    \includegraphics[scale=0.4]{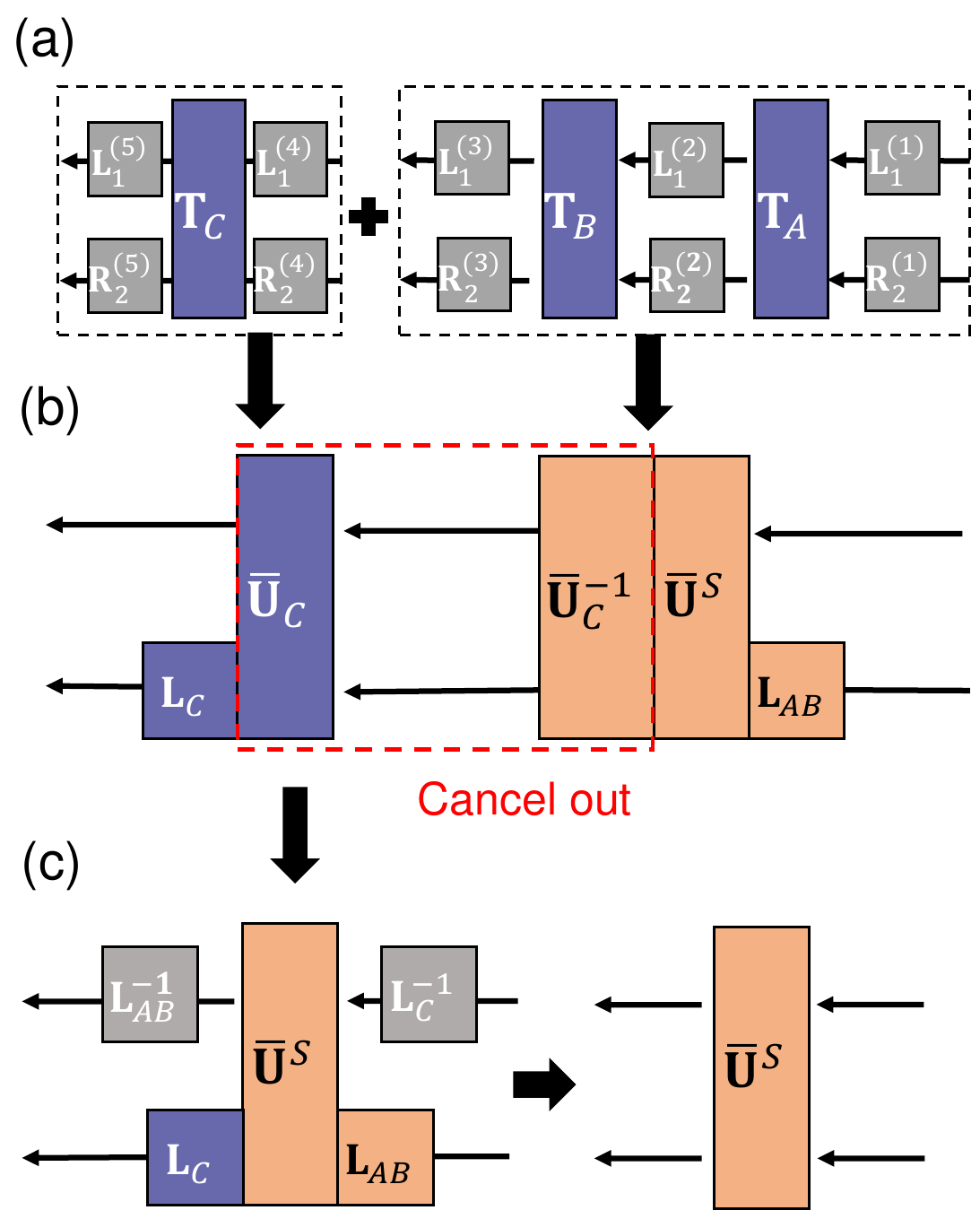}
    \caption{Three-interface protocol for engineering SWAP under the squeezing restriction. (a) Component interface $A$ is combined with $B$ to form an intermediate interface $AB$ with the transmission strength $\chi_{AB}=1-\chi_C$. (b) Such interface $AB$ can be converted to a swapped-inverse of interface $C$, i.e. ${\bf \bar{U}}_C^{-1}{\bf \bar{U}}^S$, up to irreducible mode-2 operations, ${\bf L}_{AB}$. 
    (c) After eliminating ${\bf \bar{U}}_{C}$, the irreducible mode-2 operations, ${\bf L}_{AB}$ and ${\bf L}_{C}$, can be removed by swapping operations from mode 1. Finally, the standard-form SWAP is formed.}
    \label{fig_of_swap}
\end{figure}

We first discuss the protocol for SWAP, which has no additional characteristic parameter. We realize that the protocol in Sec.~\ref{engineer_Iden_SWAP} can be applied with minor modifications to take care of the irreducible squeezing and shearing of the component interfaces. 
Our protocol is illustrated in Fig.~\ref{fig_of_swap}(a). An intermediate interface $AB$ with $\chi_{AB}=1-\chi_C$ is first constructed by the two-interface module in the last subsection.
Next, the component interface $C$ is converted into the post-squeezing form and applied after $AB$, as shown in Fig.~\ref{fig_of_swap}(b). By converting $AB$ into the swapped-inverse of $C$ up to some irreducible mode-2 operations ${\bf L}_{AB}$, i.e. ${\bf T}_{AB}={\bf \bar{U}}_{C}^{-1}{\bf \bar{U}}^S {\bf L}_{AB}$, the non-trivial interface $C$ will be cancelled. A SWAP will be remained together with mode-2 operations (Fig.~\ref{fig_of_swap}(c)), which can be eliminated by swapping squeezing and rotating from mode 1. At the end, we will have a SWAP in the standard form.

\subsection{Interfaces with $\chi\neq 0,1$}\label{sec_irreducible_squeezing}

The next protocol is for engineering an interface with $\chi\neq 0,1$. Our goal is to obtain the target interface with the transmission strength $\chi_{\text{tgt}}$ and the irreducible squeezing $\Lambda_{\text{tgt}}$. Our protocol involves four non-trivial component interfaces $A$, $B$, $C$ and $D$. The strategy is to match the characteristics parameters sequentially: two intermediate interfaces $AB$ and $CD$ are first engineered that their combined transmission strength matches $\chi_{\text{tgt}}$, then $\Lambda_{\text{tgt}}$ is matched through controlling the squeezing in between.

\subsubsection{Tuning $\chi_{ABCD}$}\label{tuning_no_Z}

\begin{figure}
    \centering
    \includegraphics[scale=0.35]{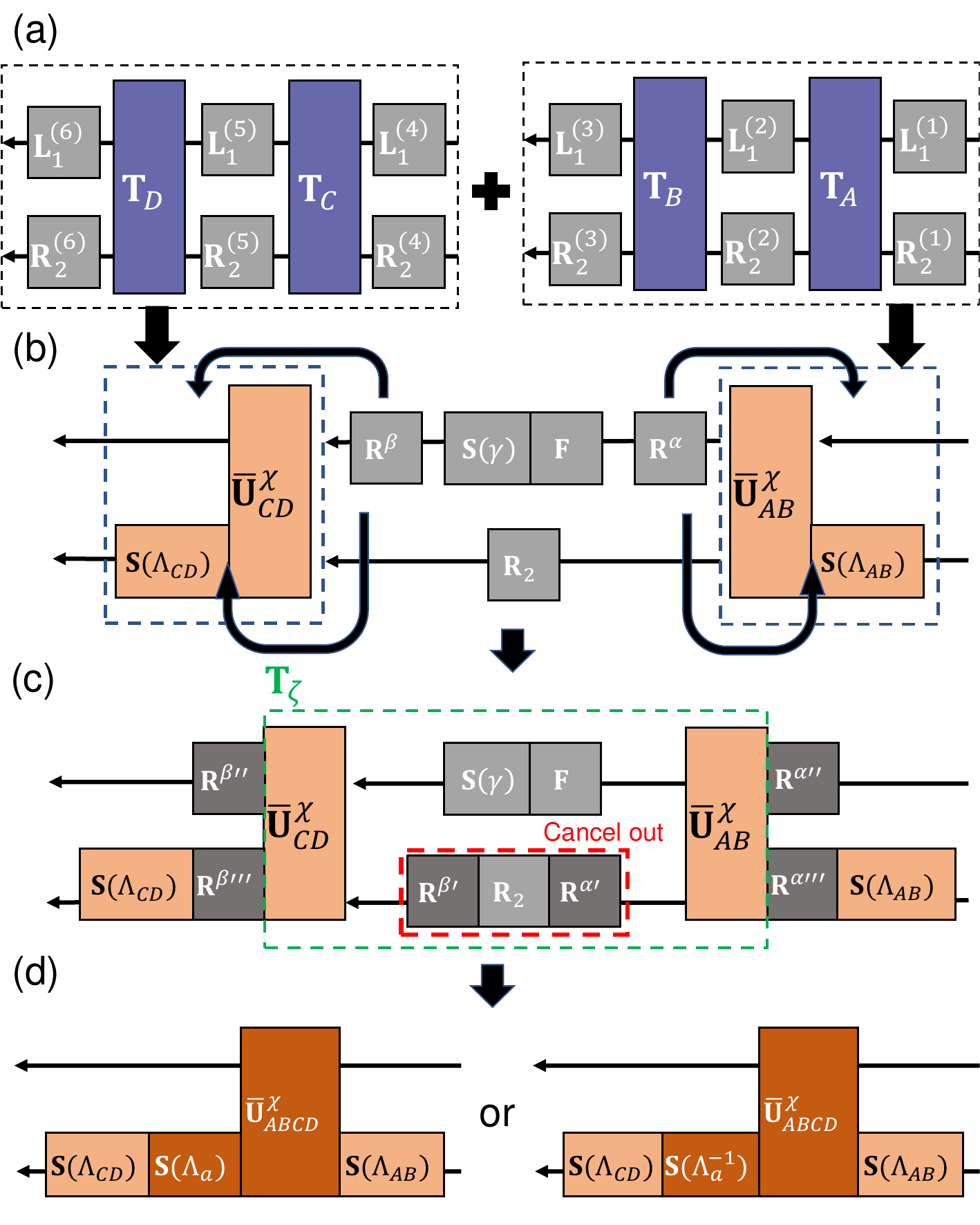}
    \caption{Four-interface protocol to engineer an interface with $\chi\neq 0,1$ under squeezing restriction.
    (a) Each pair of two component interfaces are combined to form the intermediate interfaces $AB$ and $CD$. They are then converted to pre- and post-squeezing form. (b) According to Eqs.~(\ref{effective_rotation1})-(\ref{effective_rotation2}), the controllable mode-$1$ rotations ${\bf R}_1^{\alpha}$ and ${\bf R}_1^\beta$ effectively introduces controllable mode-$2$ rotations between the standard-form transformation and the irreducible squeezing. (c) The equivalent circuit after the rearrangement. ${\bf R}_2$ is chosen to cancel all rotation between $AB$ and $CD$, while mode-1 control follows the non-interfering setup in Fig.~\ref{interference_picture}. (d) The resultant interface $ABCD$. Its transmission strength $\chi_{ABCD}$ can be engineered to the desired value according to Eq.~\eqref{chiABCD}, and the resultant irreducible squeezing $\Lambda_{ABCD}$, which is determined by $\Lambda_{AB}$, $\Lambda_{CD}$ and the controllable $\Lambda_{a}$ can be tuned to the target value according to Eq.~\eqref{lambdaABCD} or \eqref{lambdaABCD2}.}
    \label{fig_turning_lambda}
\end{figure}

To match the transmission strength, our first step is to combine $A$ with $B$ and $C$ with $D$ to form two intermediate interfaces $AB$ and $CD$ (Fig.~\ref{fig_turning_lambda}(a)) by applying the two-interface module. $\chi_{AB}$ should be chosen according to the criteria \footnote{If $\chi_A$ (or $\chi_B$) satisfies the conditions Eq.~\eqref{condition_for_chiAB} with $\chi_{AB}$ replaced by it, it is not necessary to construct interface $AB$. We can replace the role of interface $AB$ by $A$ (or $B$), and hence reduce the required amount of interfaces.} :
  \begin{eqnarray}\label{condition_for_chiAB}
  \begin{cases}
  \chi_{AB}>\chi_{\text{tgt}} & \text{for} \quad \chi_{\text{tgt}}<0, 1/2<\chi_{\text{tgt}}<1\\
  \chi_{AB}<\chi_{\text{tgt}} & \text{for} \quad 0<\chi_{\text{tgt}}\leq 1/2, \chi_{\text{tgt}}>1
  \end{cases},
  \end{eqnarray}
and $\chi_{CD}$ should be tuned according to
\begin{eqnarray}\label{chi_tot}
\chi_{AB} +\chi_{CD} -2\chi_{AB}\chi_{CD} =\chi_{\text{tgt}}.
\end{eqnarray} 
The reason for the above choice is to guarantee the combination of intermediate interfaces is either BS+BS, TMS+TMS or TMS+sTMS, otherwise the resultant irreducible squeezing $\Lambda_{ABCD}$ will be bounded (details in Appendix D). 

Secondly, we convert $AB$ and $CD$ into respectively pre- and post-squeezing forms, as illustrated in Fig.~\ref{fig_turning_lambda}(b). The mode-1 operation in between $AB$ and $CD$ is chosen as ${\bf R}^\beta_1{\bf S}_1(\gamma){\bf F}_1{\bf R}_1^\alpha$. As illustrated in Fig.~\ref{fig_turning_lambda}(c), the purpose of ${\bf R}_1^\alpha$ and ${\bf R}^\beta_1$ is to introduce a controllable rotations, ${\bf R}^{\alpha'''}_2$ and ${\bf R}^{\beta'''}_2$, between the standard-form interfaces and the irreducible mode-$2$ operations through the following equivalent circuits,
\begin{eqnarray}
{\bf R}^\alpha_1{\bf \bar{U}}^\chi_{AB} &=& {\bf R}^{\alpha'}_2 {\bf \bar{U}}^\chi_{AB} {\bf R}^{\alpha''}_1 {\bf R}^{\alpha'''}_2 ,\label{effective_rotation1}\\
{\bf \bar{U}}^\chi_{CD}{\bf R}^\beta_1 &=& {\bf R}^{\beta''}_1 {\bf R}^{\beta'''}_2 {\bf \bar{U}}^\chi_{CD} {\bf R}^{\beta'}_2,\label{effective_rotation2}
\end{eqnarray} 
where the relation between the rotations can be found in Appendix B.  
As will be discussed in Sec.~\ref{manipulating_Lambda}, this setting will be helpful to engineer the desired irreducible squeezing. Other rotations listed in Eqs.~(\ref{effective_rotation1}-\ref{effective_rotation2}) will be cancelled by the controllable rotations, such that the overall transmission strength is determined by ${\bf T}_{\zeta}\equiv  {\bf\bar{U}}^\chi_{CD}{\bf S}_1(\gamma){\bf F}_1{\bf \bar{U}}^\chi_{AB}$, which is labelled in Fig.~\ref{fig_turning_lambda}(c). 

\begin{figure}
    \centering
    \includegraphics[scale=0.8]{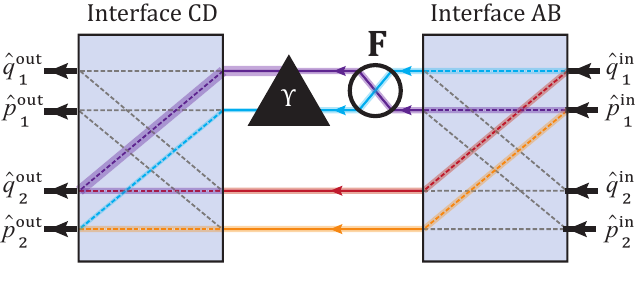}
    \caption{Illustrating the shutdown of transmission amplitude interference, i.e. $Z=0$ for any $\gamma$. After passing through the intermediate interface $AB$, a Fourier gate ${\bf F}\equiv {\bf R}(\pi/2)$ is applied on mode 1 to interchange the $q$- and $p$-quadratures. After passing through intermediate interface $CD$, each output quadrature will contain transmitted information from both input quadratures, each coming from an independent path. There is thus no interference, and the squeezing on mode 1 (triangle) does not alter the resultant transmission strength. }
    \label{interference_picture}
\end{figure}

As illustrated in Fig.~\ref{interference_picture}, the purpose of ${\bf F}_1$ is to exchange the $q$- and $p$-quadratures of mode 1 after passing through $AB$, so the quadrature interference between $AB$ and $CD$ can be switched off. This can be seen from the transmission matrix of ${\bf T}_{\zeta}$,
\begin{eqnarray}\label{turnoffZ}
{\bf T}^{21}_{\zeta} 
&=&\begin{pmatrix}
\bar{U}_{CD}^{22,q}\bar{U}_{AB}^{21,q} & -\gamma^{-1} \bar{U}_{CD}^{21,q} \bar{U}_{AB}^{11,p}\\
\gamma \bar{U}_{CD}^{21,p} \bar{U}_{AB}^{11,q} & \bar{U}_{CD}^{22,p} \bar{U}_{AB}^{21,p}
\end{pmatrix},
\end{eqnarray}
where ${\bf \bar{U}}_{\alpha}^{ij}\equiv \textrm{diag}(\bar{U}_{\alpha}^{ij,q}, \bar{U}_{\alpha}^{ij,p})$ for $\alpha=AB,CD$.
The amplification and deamplification due to the controllable mode-1 squeezing ${\bf S}_1(\gamma)$ appear only in the off-diagonal entries of Eq.~\eqref{turnoffZ}, so their effects in the transmission strength compensate each other, i.e. $\chi_{ABCD} \equiv \det({\bf T}^{21}_{\zeta})$ is independent of $\gamma$. 
We find that the overall transmission strength is then given by \begin{eqnarray}
\chi_{ABCD}=\chi_{AB} + \chi_{CD}-2\chi_{AB}\chi_{CD},\label{chiABCD}
\end{eqnarray}
which is determined by the strengths of $AB$ and $CD$ only. Then, the choice of $\chi_{AB}$ and $\chi_{CD}$ guarantees that $\chi_{ABCD}$ matches $\chi_{\text{tgt}}$  (c.f. Eq.~(\ref{chi_tot})).

 \subsubsection{Manipulating $\Lambda_{ABCD}$}\label{manipulating_Lambda}

Although the resultant transmission strength is independent of the in-between mode-$1$ squeezing ${\bf S}_1(\gamma)$, the resultant irreducible squeezing does depend on $\gamma$.  This gives us the ability to engineer the resultant interface with the desired irreducible squeezing, i.e. $\Lambda_{ABCD}=\Lambda_{\text{tgt}}$. 
Explicitly, thanks to the controllable rotations ${\bf R}^{\alpha'''}_2$  and ${\bf R}^{\beta'''}_2$ introduced by the equivalent circuit in Eqs.~(\ref{effective_rotation1})-(\ref{effective_rotation2}), the intermediate interface ${\bf T}_{\zeta}$ can be quadrature-diagonalized, as illustrated in Fig.~\ref{fig_turning_lambda}(d). The resultant irreducible squeezing is then given by a simple product of those of the intermediate interfaces $AB$, $CD$, and ${\bf T}_{\zeta}$, i.e. 
\begin{equation}\label{lambdaABCD}
    \Lambda_{ABCD}=\Lambda_{AB}\Lambda_{CD}\Lambda_{a},
\end{equation}
or the division of them, i.e.,
\begin{equation}\label{lambdaABCD2}
    \Lambda_{ABCD}=\Lambda_{AB}\Lambda_{CD}/\Lambda_{a}.
\end{equation}
$\Lambda_{AB}$ and $\Lambda_{CD}$ are by-products of engineering $AB$ and $CD$ and assumed to be untunable; $\Lambda_{a} \geq 1$ is determined by the singular values of ${\bf T}_{\zeta}^{22}$, which is controllable by the mode-1 squeezing $\gamma$.  In Appendix D, we show that $\Lambda_a$, and hence $\Lambda_{ABCD}$, can be tuned to arbitrary values. Overall, this protocol can engineer an interface with arbitrary  $\chi_{\text{tgt}} \neq 0,1$ and $\Lambda_{\text{tgt}}$.

\subsection{sQNDIs}\label{SQND}

\begin{figure}
    \centering
    \includegraphics[scale=0.45]{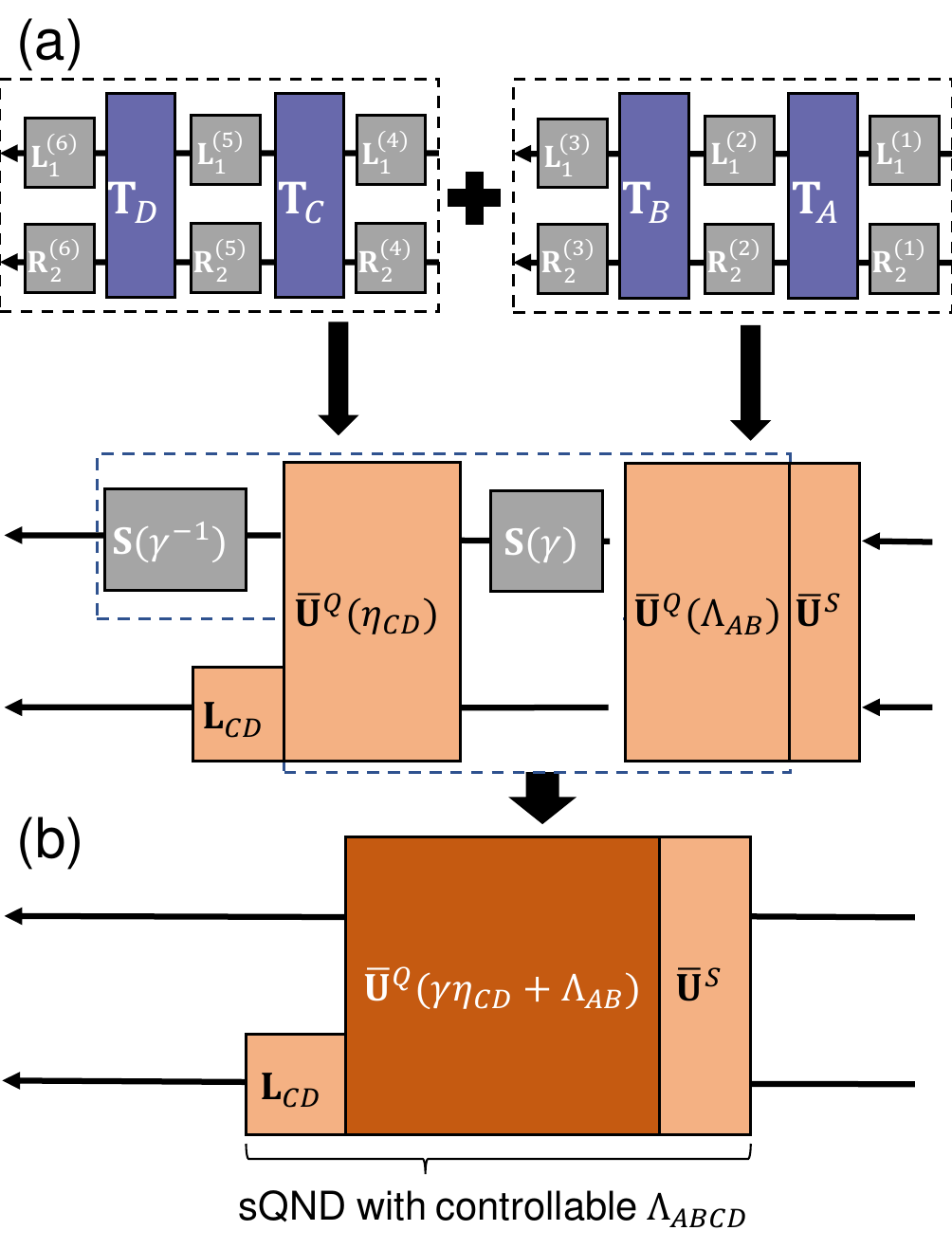}
    \caption{Four-interface protocol for constructing arbitrary sQND under squeezing restriction. (a) Component interfaces $A$ and $B$ are combined to form a sQNDI while $C$ and $D$ are forming a QNDI. The QND strength of the ${\bf \bar{U}}^Q(\eta_{CD})$ is modified by controllable mode-1 squeezing before and afterwards according to Eq.~\eqref{QND_squeezing}. Subsequently, it is combined with ${\bf \bar{U}}^Q(\Lambda_{AB})$ to form a standard-form QNDI with the controllable strength $\gamma \eta_{CD}+\Lambda_{AB}$. (b) The resultant interface is a sQNDI. Its QND strength, which is determined by both the controllable strength $\gamma \eta_{CD}+\Lambda_{AB}$ and the irreducible squeezing and shearing of $CD$, is given by Eq.~\eqref{lambda_sQND}. }
    \label{fig_SQND}
\end{figure}

Our protocol to engineer arbitrary sQNDIs involves four non-trivial component interfaces. As illustrated in Fig.~\ref{fig_SQND}(a), by using the two-interface module in Sec.~\ref{two_interface_module}, we first combine the component interfaces $A$ and $B$ to form a standard-form sQNDI, and similarly $C$ with $D$ to form a QNDI in the post-squeezing form. 
Next, the QND strength of $CD$ is manipulated by using the mode-1 squeezing according to Eq.~\eqref{QND_squeezing}.
Finally, by recognizing that a sQNDI is equivalent to a QNDI followed by a SWAP, i.e. ${\bf \bar{U}}^{SQ}(\Lambda) ={\bf \bar{U}}^{Q}(\Lambda){\bf \bar{U}}^S$, and applying two standard-form QNDIs in sequence will result in a QNDI with the sum of their strengths, i.e. ${\bf \bar{U}}^Q(\eta){\bf \bar{U}}^{Q}(\eta')={\bf \bar{U}}^{Q}(\eta+\eta')$,
cascading $AB$ with $CD$ will result in a sQNDI with a controllable QND strength, and equivalently a controllable irreducible squeezing.
By considering the explicit expression of the mode-2 reflection matrix, including the irreducible squeezing and shearing of $CD$, we show in Appendix D that the resultant irreducible squeezing strength is given by
\begin{eqnarray}
\Lambda_{ABCD} &=& \left(\gamma \eta_{CD} +\Lambda_{AB}\right)
\nonumber\\
&\times&\frac{\sqrt{(\Lambda_{CD}^4+1)+(\Lambda_{CD}^4-1)\cos\phi_{CD}}}{2\Lambda_{CD}},\label{lambda_sQND}
\end{eqnarray}
where $\tan\phi_{CD} \equiv \kappa_{CD} \Lambda^2_{CD}$. By controlling $\gamma$, $\Lambda_{ABCD}$ can be tuned to any desired value.

\subsection{QNDIs}\label{QND}

Surprisingly, any QNDI can be engineered by using also four non-trivial interfaces, even though there is one more parameter to be matched. Our strategy is to reverse engineer the required single-mode controls by using the sQND engineering protocol in Sec.~\ref{SQND}. Explicitly, we first construct two standard-form sQNDIs by combining the component interfaces $A$ with $B$ and $C$ with $D$ according to the two-interface-module. Our aim is to explore the single-mode controls, ${\bf L}^{(n)} (n=1,2,3)$, that can generate the desired QNDI when cascading with the two intermediate sQNDI components, i.e.
\begin{eqnarray}\label{inverse_1}
&&{\bf L}^{(3)}{\bf \bar{U}}^{SQ}(\Lambda_{CD}) {\bf L}^{(2)} {\bf \bar{U}}^{SQ}(\Lambda_{AB}) {\bf L}^{(1)} \nonumber\\
&=& {\bf \bar{U}}^{Q}(\eta){\bf R}_2(\phi_{\text{tgt}}){\bf S}_2(\Lambda_{\text{tgt}}),
\end{eqnarray}
where $\kappa_{\text{tgt}} \equiv \tan\phi_{\text{tgt}}$ and $\Lambda_{\text{tgt}}$ are respectively the target irreducible shearing and squeezing. We note that the QND strength $\eta$ is unimportant as it is adjustable by mode-$1$ operation. 

By rewriting Eq.~\eqref{inverse_1} as
\begin{eqnarray}\label{inverse_2}
&&{\bf L}^{(1)} \left({\bf \bar{U}}^Q(\eta){\bf R}_2(\phi_{\text{tgt}}){\bf S}_2(\Lambda_{\text{tgt}})\right)^{-1}{\bf L}^{(3)}{\bf \bar{U}}^{SQ}(\Lambda_{CD}){\bf L}^{(2)} \nonumber\\
&=& \left({\bf \bar{U}}^{SQ}(\Lambda_{AB})\right)^{-1},
\end{eqnarray}
it coincides with the configuration of sQNDI engineering in Sec.~\ref{SQND}, i.e. a sQNDI with strength $\Lambda_{AB}$ is constructed by cascading a sQNDI with strength $\Lambda_{CD}$ and a QNDI with irreducible shearing $\kappa_{\text{tgt}}$ and squeezing $\Lambda_{\text{tgt}}$. By applying our protocol in Sec.~\ref{SQND}, the required ${\bf L}^{(n)}$ can be identified. 

\subsection{Remote squeezing}\label{remote_gate}

Finally, we discuss the protocol to engineer an arbitrary Identity class interface, i.e. an interface with $n_T=0$, $\chi=0$ and a controllable $\Lambda$. Under the squeezing restriction, a general interface in this class is no longer an Identity operation but an irreducible squeezing in mode 2. Constructing this interface is thus equivalent to "remotely" squeeze the restricted mode with any desired strength $\Lambda_{\text{tgt}}$ through interfacing with an active mode; we thus call this type of protocol \textit{remote squeezing}. 
Remote squeezing can be straightforwardly implemented by using two rounds of SWAP. After the first SWAP, the mode-$2$ initial state is transferred to mode 1 and then directly squeezed. With the second SWAP, the squeezed initial state is sent back to mode 2. By using the three-interface SWAP engineering protocol in Sec.~\ref{swap_engineering_restriction}, this double-swap method can be implemented with any six non-trivial component interfaces. 

\begin{figure}
    \centering
    \includegraphics[scale=0.32]{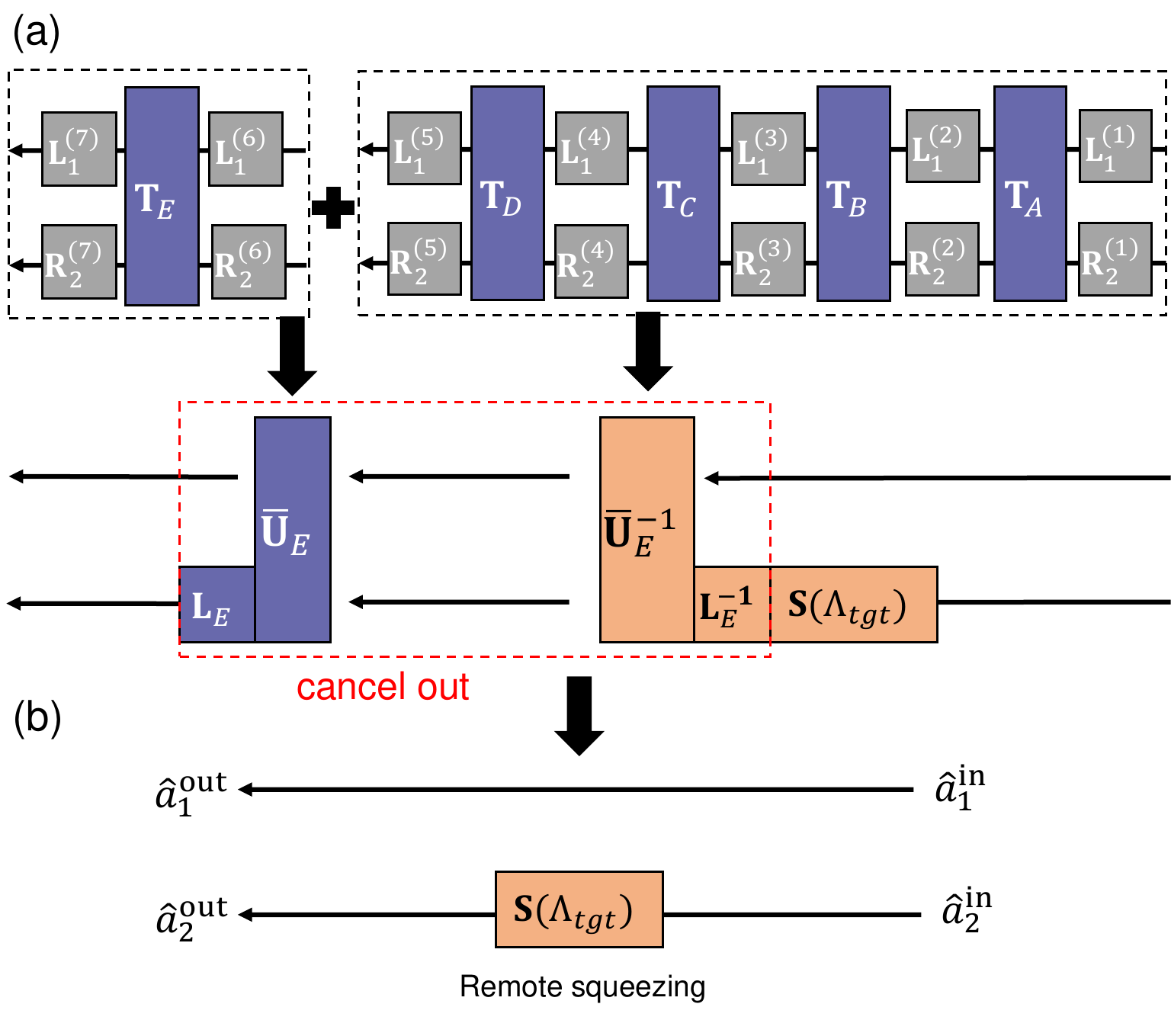}
    \caption{Five-interface protocol for engineering remote squeezing. (a) By using the four-interface protocol for engineering arbitrary non-trivial interface, component interfaces $A, B, C, D$ can always be combined to become the inverse of the component $E$ preceded with an extra mode-2 squeezing ${\bf S}(\Lambda_{\text{tgt}})$. (b) Cascading these interfaces will result in a remote squeezing with strength $\Lambda_{\text{tgt}}$.}
    \label{fig_of_remote_squeezing}
\end{figure}

Moreover, we developed a simplified remote squeezing scheme that requires only five non-trivial component interfaces, as illustrated in Fig.~\ref{fig_of_remote_squeezing}. 
Following the four-interface protocols in Sec.~\ref{sec_irreducible_squeezing}-\ref{QND}, our idea is to use the first four components to engineer an intermediate interface $ABCD$ that is the inverse of the fifth component $E$ preceded with the desired mode-$2$ squeezing, i.e. ${\bf T}_{ABCD}={\bf T}_E^{-1}{\bf S}_2(\Lambda_{\text{tgt}})$. After cascading with $E$, the non-trivial transformations are eliminated and the resultant interface becomes a remote squeezing at our desired value, ${\bf S}_2(\Lambda_{\text{tgt}})$.

As a remark, we have also presented other remote squeezing schemes in Appendix E that could use fewer component interfaces except a few special cases.

\section{Conclusion}\label{conclusion}

We have completely characterized the linear bosonic two-mode interfaces under single-mode operational constraints. We have studied two situations: the general situation that any single-mode operation is available on both modes, and the restricted situation that squeezing is available on one mode only. We recognized the classifiers that are useful in identifying interfaces: interfaces with different classifiers are not inter-convertible by any allowable single-mode control. Under the general situation, any non-trivial interface can be classified by its transmission strength $\chi$. For the restricted situation, we discovered two additional invariant classifiers: irreducible squeezing $\Lambda$ and irreducible shearing $\kappa$.  The main results of characterization are listed in Table.~\ref{main_table}. 

Guided by the characterization, we developed the protocols for engineering arbitrary linear two-mode interfaces with any available interface in the platform. Our protocols incorporate the available interfaces as components, and cascade multiple of them with suitable single-mode controls. 
For the general situation, our protocol can construct any non-trivial interface by using only two rounds of component interface, and SWAP and Identity by using only three rounds. We prove that our protocols are optimal because they involve the fewest possible rounds of component interface. Under the squeezing restriction, we also developed the modified protocols to engineer the additional invariant classifiers. The required number of component interface remains three for engineering SWAP, while four is needed for engineering other interfaces. To resolve the squeezing restriction, we introduce the remote squeezing protocol to squeeze a restricted mode through interfacing with an active mode.
We summarize the number of required interfaces in Table~\ref{main_table}. 

Our characterization can help benchmarking experiment platforms and optimizing their designs for implementing specific applications.
Our interface engineering protocols can overcome the limitations of physical platforms in implementing interfaces. They thus facilitate the realization of quantum technologies in a wider range of platforms, such as implementing universal CV logic gates in passive systems, reducing noise in quantum transduction, and conducting interferometry between hybrid quantum systems.
Moreover, our remote squeezing scheme can help generating resourceful states for sensing, communication and computation, such as spin squeezed state \cite{squeezed_spin_state1993} and squeezed optical state \cite{squeezed_1983}.  

\begin{acknowledgments}
This work is supported by the Natural Sciences and Engineering Research Council of Canada (NSERC RGPIN-2021-02637) and Canada Research Chairs (CRC-2020-00134).
Sheung Chi thanks the Summer Undergraduate Research Exchange (SURE) Program by CUHK for providing valuable research opportunities and support.
\end{acknowledgments}

\appendix

\section{Matrix form of single-mode operations}

We list the expression of the single-mode operations, especially the rotation and squeezing, in this appendix. In the main text, we denote mode-1 and mode-2 operations by the $4 \times 4$ matrices ${\bf L}_1$ and ${\bf L}_2$ respectively. Since ${\bf L}_1$ and ${\bf L}_2$ transform only one mode, their off-diagonal blocks should be null and the block acting on the other mode should be identity, i.e.
\begin{eqnarray}
 {\bf L}_1 \equiv\begin{pmatrix}
{\bf L}_1^{11} & {\bf 0} \\
{\bf 0} & {\bf I} 
\end{pmatrix} \quad \text{and}\quad
{\bf L}_2 \equiv \begin{pmatrix}
{\bf I} & {\bf 0} \\
{\bf 0} & {\bf L}_2^{22} 
\end{pmatrix}.
\end{eqnarray}

Single-mode rotation induces phase shift to the mode operator, i.e. $\hat{a}_j^{\text{out}} = e^{-i\phi}\hat{a}_j^{\text{in}}$ or
\begin{eqnarray}
\begin{pmatrix}
\hat{q}_j^{\text{out}}\\
\hat{p}_j^{\text{out}}
\end{pmatrix} =  \begin{pmatrix}
\cos\phi & -\sin\phi \\
\sin\phi & \cos\phi 
\end{pmatrix}
\begin{pmatrix}
\hat{q}_j^{\text{in}}\\
\hat{p}_j^{\text{in}}
\end{pmatrix}
\end{eqnarray}
for $j=1,2$.
Therefore, the mode-$j$ rotation operation denoted by ${\bf R}_j(\phi)$ should have the non-identity block
\begin{eqnarray}\label{r_phi}
[{\bf R}_j(\phi)]^{jj} =
\begin{pmatrix}
\cos\phi & -\sin\phi \\
\sin\phi & \cos\phi 
\end{pmatrix}.
\end{eqnarray}

Single-mode squeezing amplifies one quadrature while deamplifies the other, i.e. it transforms the quadrature operators as
\begin{eqnarray}
\begin{pmatrix}
\hat{q}_j^{\text{out}}\\
\hat{p}_j^{\text{out}}
\end{pmatrix} =  \begin{pmatrix}
\gamma & 0 \\
0 & \gamma^{-1} 
\end{pmatrix}
\begin{pmatrix}
\hat{q}_j^{\text{in}}\\
\hat{p}_j^{\text{in}}
\end{pmatrix}.
\end{eqnarray}
Therefore, the mode-$j$ squeezing operation denoted by ${\bf S}_j(\gamma)$ should satisfy
\begin{eqnarray}\label{s_gamma}
[{\bf S}_j(\gamma)]^{jj} = \begin{pmatrix}
\gamma & 0 \\
0 & \gamma^{-1} 
\end{pmatrix}.
\end{eqnarray}

\section{Resultant interfaces of two-interface setup}

\subsection{Necessary and sufficient conditions for engineering Identity and SWAP}

In this appendix, we will show that it is impossible to construct Identity nor SWAP by the two-interface setup, $ {\bf \bar{U}}_B {\bf L}_1 {\bf L}_2{\bf \bar{U}}_A$, unless the component interfaces satisfying some specific conditions. 
We will exploit all the possible combinations of component interfaces and single-mode operations ${\bf L}_1$ and ${\bf L}_2$. 

First, we recognize some combinations of single-mode operations are redundant.
Each Gaussian single-mode operation is generally characterized by 3 parameters since it can always be decomposed into a rotation-squeezing-rotation sequence \cite{rmp2012}, i.e. 
${\bf L}_i = {\bf R}_i^{(2)} {\bf S}_i {\bf R}_i^{(1)}.$ Then, the two-interface setup contains 6 single-mode operational parameters, however not all parameters are independent. To identify the independent parameters, we rearrange the single-mode operations and eliminate the redundant parameters. The rearrangement is based on the fact that the standard-form interface is invariant under some single-mode transformations, i.e.
\begin{eqnarray}\label{exchange}
{\bf L}_1^x {\bf L}_2^y {\bf \bar{U}} {\bf L}_1^u{\bf L}_2^v \equiv {\bf \bar{U}},
\end{eqnarray}
where ${\bf L}_1^u$, ${\bf L}_2^v$, ${\bf L}_1^x$ and ${\bf L}_2^y$ will be determined later. By applying Eq.~\eqref{exchange} on both interfaces, we can rewrite the two-interface setup as
\begin{eqnarray}\label{exchange2}
&&{\bf \bar{U}}_B {\bf L}_1 {\bf L}_2{\bf \bar{U}}_A \nonumber\\
&=& ({\bf L}_1^{B,x})^{-1} ({\bf L}^{B,y}_{2})^{-1} {\bf \bar{U}}_B {\bf L}_1'{\bf L}_2' {\bf \bar{U}}_A ({\bf L}_1^{A,u})^{-1} ({\bf L}^{A,v}_{2})^{-1},\nonumber\\
\end{eqnarray}
where 
\begin{eqnarray}
{\bf L}_1' &=& ({\bf L}_1^{B,u})^{-1} {\bf L}_1 ({\bf L}_1^{A,x})^{-1},\label{L1ppp}\\
{\bf L}_2' &=&({\bf L}_2^{B,v})^{-1}{\bf L}_2({\bf L}_2^{A,y})^{-1},\label{L2ppp}
\end{eqnarray}
will be shown to be characterized by fewer parameters. 
Since the single-mode transformations after interface $B$, ${\bf L}_1^{B,x}$ and ${\bf L}_2^{B,y}$, and that before interface $A$, ${\bf L}_1^{A,u}$ and ${\bf L}_2^{A,v}$, do not alter the class and transmission strength of the resultant interface $AB$, we can examine the simplified configuration, ${\bf T}_{AB}\equiv {\bf \bar{U}}_B {\bf L}_1'{\bf L}_2' {\bf \bar{U}}_A$, for the transmission strength $\chi_{AB} \equiv \text{det}({\bf T}_{AB}^{21})$. 

To explore the single-mode transformations satisfying Eq.~\eqref{exchange}, we rewrite Eq.~\eqref{exchange} as
\begin{eqnarray}
{\bf \bar{U}} {\bf L}_1^u {\bf L}_2^v ({\bf \bar{U}})^{-1} = ({\bf L}_1^x)^{-1} ({\bf L}_2^y)^{-1}, \label{LxLy}
\end{eqnarray}
that is explicitly
\begin{eqnarray}\label{matrix_condition}
&&\begin{pmatrix}
{\bf \bar{U}}^{11} & {\bf \bar{U}}^{12} \\
{\bf \bar{U}}^{21} & {\bf \bar{U}}^{22} 
\end{pmatrix}
 \begin{pmatrix}
[{\bf L}_1^u]^{11} & {\bf 0} \\
{\bf 0} & [{\bf L}^v_2]^{22} 
\end{pmatrix}
 \begin{pmatrix}
{\bf \bar{U}}^{22} & -{\bf \bar{U}}^{12} \\
-{\bf \bar{U}}^{21} & {\bf \bar{U}}^{11} 
\end{pmatrix}
\nonumber\\
&=&\begin{pmatrix}
([{\bf L}_1^x]^{11})^{-1} & {\bf 0} \\
{\bf 0} & ([{\bf L}_2^y]^{22})^{-1} 
\end{pmatrix}.\label{block_matrix} 
\end{eqnarray}
By considering the off-diagonal block matrices of Eq.~\eqref{matrix_condition}, we obtain the conditions of ${\bf L}_1^u$ and ${\bf L}_2^v$ that are given by
\begin{eqnarray}
-{\bf \bar{U}}^{11} [{\bf L}^u_1]^{11} {\bf \bar{U}}^{12} + {\bf \bar{U}}^{12} [{\bf L}_2^v]^{22} {\bf \bar{U}}^{11} &=& {\bf 0}, \label{Lq_condition1}\\
{\bf \bar{U}}^{21} [{\bf L}_1^u]^{11} {\bf \bar{U}}^{22} - {\bf \bar{U}}^{22} [{\bf L}^v_2]^{22} {\bf \bar{U}}^{21}  &=& {\bf 0}.\label{Lq_condition2}
\end{eqnarray}
By considering the explicit form of ${\bf \bar{U}}$, we find that Eqs.~\eqref{Lq_condition1} and \eqref{Lq_condition2} are equivalent. We have two unknowns, ${\bf L}_1^u$ and ${\bf L}_2^v$, but one equation, it implies that there exists a family of ${\bf L}^u_1$ and ${\bf L}^v_2$ satisfying the invariant condition Eq.~\eqref{exchange} for each class of interface.

When the interface is BS, TMS or sTMS, since their ${\bf \bar{U}}^{ij}$'s are of rank 2, there is no restriction to the matrix ${\bf L}_2^v$ and we can choose any mode-2 operation. Then, ${\bf L}^u_1$ is determined by Eq.~\eqref{Lq_condition1} or \eqref{Lq_condition2}. After some algebras, we obtain
\begin{eqnarray}\label{M_resolve}
[{\bf L}^u_1]^{11}
 &=&  \begin{pmatrix}
L_{11}^v & \pm L_{12}^v \\
\pm L_{21}^v & L_{22}^v 
\end{pmatrix},
\end{eqnarray}
where $L^v_{ij}$ is the matrix element of the block matrix $[{\bf L}^v_2]^{22}$, $+$ sign is for BS and $-$ sign is for TMS and sTMS. 
Then, ${\bf L}^x_1$ and ${\bf L}^y_2$ are determined by the diagonal block matrices of Eq.~\eqref{matrix_condition}, 
\begin{eqnarray}
{\bf \bar{U}}^{11} [{\bf L}^u_1]^{11} {\bf \bar{U}}^{22} -{\bf \bar{U}}^{12} [{\bf L}^v_2]^{22} {\bf \bar{U}}^{21} &=& ([{\bf L}^x_1]^{11})^{-1} ,\label{LxLpLq}\\
-{\bf \bar{U}}^{21} [{\bf L}^u_1]^{11} {\bf \bar{U}}^{12} + {\bf \bar{U}}^{22} [{\bf L}^v_2]^{22} {\bf \bar{U}}^{11} &=& ([{\bf L}^y_2]^{22})^{-1} .\label{LyLpLq}
\end{eqnarray}
By substituting Eq.~\eqref{M_resolve} into Eqs.~\eqref{LxLpLq} and \eqref{LyLpLq}, we obtain
\begin{eqnarray}
[{\bf L}^x_1]^{11} &=& \begin{cases}
([{\bf L}_1^u]^{11})^{-1}  & \text{for BS, TMS}\\
([{\bf L}^v_2]^{22})^{-1}  &\text{for sTMS}
\end{cases},\label{Lx}\\
{[\bf L}^y_2]^{22} &=& \begin{cases}
([{\bf L}^u_1]^{11})^{-1}  & \text{for BS, TMS}\\
([{\bf L}^v_2]^{22})^{-1}  &\text{for sTMS}
\end{cases}.\label{Ly}
\end{eqnarray}

When the interface is QNDI (sQNDI), since its transmission (reflection) matrices are of rank $1$, only the upper triangular $[{\bf L}^u_1]^{11}$ and  lower triangular $[{\bf L}^v_2]^{22}$, i.e.  
\begin{eqnarray}
[{\bf L}^u_1]^{11} = \begin{pmatrix}
a^{-1} & b \\
0 & a 
\end{pmatrix},
[{\bf L}^v_2]^{22} = \begin{pmatrix}
a & 0 \\
c & a^{-1} 
\end{pmatrix},\label{LpQND}
\end{eqnarray}
satisfy Eqs.~(\ref{Lq_condition1}) - (\ref{Lq_condition2}), where $a, b, c$ are arbitrary constants. Moreover, we can determine ${\bf L}^x_1$ and ${\bf L}^y_2$ by substituting Eq.~(\ref{LpQND}) into Eq.~\eqref{block_matrix} and they are 
\begin{eqnarray}
[{\bf L}^x_1]^{11} = \begin{cases}
([{\bf L}_1^u]^{11})^{-1} & \text{for QNDI}\\
([{\bf L}_2^v]^{22})^{-1} &\text{for sQNDI}
\end{cases},\label{LxQ}\\
{[\bf L}^y_2]^{22} =\begin{cases}
([{\bf L}^u_1]^{11})^{-1} & \text{for QNDI}\\
([{\bf L}^v_2]^{22})^{-1} &\text{for sQNDI}
\end{cases}.\label{LyQ}
\end{eqnarray}

Now, we are able to determine ${\bf L}_1'$ and ${\bf L}_2'$ for different combination of component interfaces. As discussed in Sec.~\ref{combination_possibility}, it is legitimate to assume component interfaces $A$ and $B$ do not belong to Identity and SWAP, and the combination involving sTMS or sQNDI components can be deduced from the combinations of TMS or QNDI respectively, so we need to consider only six combinations:  BS+BS, TMS+TMS, BS+TMS, BS+QNDI, TMS+QNDI and QNDI+QNDI.

For BS+BS, TMS+TMS and BS+TMS, we are always able to find the suitable ${\bf L}_1^{B,u}$, ${\bf L}_2^{B,v}$, ${\bf L}_1^{A,x}$ and ${\bf L}_2^{A,y}$ to simplified ${\bf L}_1'$ as ${\bf R}_1(\phi_1){\bf S}_1(\gamma)$ and ${\bf L}_2'$ as ${\bf I}_2$ respectively according to Eqs.~\eqref{L1ppp} and \eqref{L2ppp}. Therefore, these combinations have only 2 independent single-mode operational parameters, and the corresponding simplified configuration is given by
\begin{equation}\label{config_A}
  {\bf T}_{AB} = {\bf \bar{U}}_B {\bf R}_1(\phi_1){\bf S}_1(\gamma) {\bf \bar{U}}_A.
\end{equation}

For BS+QNDI or TMS+QNDI, the rearrangement of single-mode transformations around the QNDI, which is the interface $B$, is restricted according to Eqs.~\eqref{LpQND}, \eqref{LxQ} and \eqref{LyQ}. We find that ${\bf L}_2'$ can still be chosen as identity by setting ${\bf L}_2^{B,v}={\bf I}$ and ${\bf L}_2^{A,y}= {\bf L}_2$. However, under this setting, ${\bf L}_1'$ cannot be simplified, i.e. it is still characterized by the general rotation-squeezing-rotation sequence, ${\bf L}_1'={\bf R}_1(\phi_1){\bf S}_1(\gamma){\bf R}_1(\epsilon)$. Then, the simplified configuration is given by
\begin{equation}\label{config_B}
{\bf T}_{AB} = {\bf \bar{U}}_B {\bf R}_1(\phi_1){\bf S}_1(\gamma){\bf R}_1(\epsilon){\bf \bar{U}}_A,
\end{equation}
that contains 3 independent single-mode operational parameters.

For QNDI+QNDI, we discover that even though ${\bf L}_2^{B,v}$ and ${\bf L}_2^{A,y}$ are restricted, ${\bf L}_2'$ can still be reduced to ${\bf R}_2(\phi_2)$. Explicitly, we can choose ${\bf L}_2^{B,v}={\bf I}_2$, but ${\bf R}_2(\phi_2){\bf L}_2^{A,y}  =  {\bf L}_2$. This choice is based on the QR decomposition \cite{qu_decomp}, i.e. any real matrix, $[{\bf L}_2]^{22}$, can be decomposed into an orthogonal matrix, $[{\bf R}(\phi_2)]^{22}$, and a triangular matrix, $[{\bf L}^{A,y}_2]^{22}$. Similar as BS+QNDI or TMS+QNDI, ${\bf L}_1'$
cannot be simplified, therefore the most simplified configuration is given by
\begin{equation}\label{config_C}
{\bf T}_{AB} = {\bf \bar{U}}_B {\bf R}_1(\phi_1){\bf S}_1(\gamma) {\bf R}_1(\epsilon) {\bf R}_2(\phi_2) {\bf \bar{U}}_A,
\end{equation}
that contains 4 independent single-mode operational parameters.

With the above simplified configurations, we can determine the sufficient and necessary conditions for engineering Identity and SWAP by brute force calculation based on Eqs.~(\ref{config_A})-(\ref{config_C}).
To engineer Identity interfaces, it requires the overall transmission matrix to be null. We list the transmission matrices for all combinations in Table.~\ref{transmission_matrix}. By direct calculation, we determine that a null transmission matrix can be generated if and only if two interfaces are of the same type with the identical transmission strength, i.e. $\chi_A = \chi_B$. There are three possible combinations, BS+BS, TMS+TMS and QNDI+QNDI. Specifically, for BS+BS, two BS angles are required to be equal, i.e. $|\theta_A| = |\theta_B|$, and hence two modes are split by the first interface but recombined by the second interface. For TMS+TMS, two TMS strength are equal, $|r_1| = |r_2|$, so the modes are amplified and then deamplified sequentially. 
For QNDI+QNDI, the QND strength of one QNDI is manipulated by applying single-mode squeezing before and after the interface as discussed in Sec.~\ref{classification_begin}, such that the two QNDIs will have QND strengths with the same magnitude but opposite sign. Then, the two QNDIs can cancel each other.

\begin{widetext}
\begin{table*}
    \centering
       \begin{tabular}{|c|c|}
       \hline
          Combination & Transmission matrix ${\bf T}^{AB}_{21}$\\
    \hline
    BS+BS   &  $
        \begin{pmatrix}
    -\cos\theta_B \sin\theta_A -\sin\theta_B\cos\theta_A\gamma\cos\phi_1 & \sin\theta_B\cos\theta_A \gamma^{-1}\sin\phi_1\\
    -\sin\theta_B\cos\theta_A\gamma\sin\phi_1 & -\cos\theta_B \sin\theta_A - \sin\theta_B\cos\theta_A\gamma^{-1}\cos\phi_1
    \end{pmatrix}$\\\hline
    TMS+TMS & $\begin{pmatrix}
    \cosh r_B \sinh r_A +\sinh r_B\cosh r_A\gamma\cos\phi_1 & -\sinh r_B\cosh r_A \gamma^{-1}\sin\phi_1\\
    -\sinh r_B\cosh r_A\gamma\sin\phi_1 & -\cosh r_B \sinh r_A - \sinh r_B\cosh r_A\gamma^{-1}\cos\phi_1
    \end{pmatrix}$\\\hline
    BS+TMS   & $\begin{pmatrix}
    \cos\theta_B \sinh r_A -\sin\theta_B\cosh r_A\gamma\cos\phi_1 & \sin\theta_B\cosh r_A \gamma^{-1}\sin\phi_1\\
    -\sin\theta_B\cosh r_A\gamma\sin\phi_1 & -\cos\theta_B \sinh r_A - \sin\theta_B\cosh r_A\gamma^{-1}\cos\phi_1
    \end{pmatrix}$\\\hline
    BS+QNDI & $\begin{pmatrix}
\sin\theta_B\eta_A -\cos\theta_B \left(\gamma\cos\epsilon\cos\phi_1-\gamma^{-1}\sin\epsilon\sin\phi_1\right)& \sin\theta_B\left(\gamma\cos\phi_1\sin\epsilon+\gamma^{-1}\cos\epsilon\sin\phi_1\right)\\
-\sin\theta_B\left(\gamma\cos\epsilon\sin\phi_1 + \gamma^{-1}\cos\phi_1\sin\epsilon\right)& -\sin\theta_B \left(\gamma^{-1}\cos\epsilon\cos\phi_1-\gamma\sin\epsilon\sin\phi_1\right)
    \end{pmatrix}$\\\hline
    TMS+QNDI &  $\begin{pmatrix}
\sinh r_B\eta_A + \cosh r_B \left(\gamma\cos\epsilon\cos\phi_1-\gamma^{-1}\sin\epsilon\sin\phi_1\right)& -\sinh r_B\left(\gamma\cos\phi_1\sin\epsilon+\gamma^{-1}\cos\epsilon\sin\phi_1\right)\\
-\sinh r_B\left(\gamma\cos\epsilon\sin\phi_1 + \gamma^{-1}\cos\phi_1\sin\epsilon\right)& -\sinh r_B \left(\gamma^{-1}\cos\epsilon\cos\phi_1-\gamma\sin\epsilon\sin\phi_1\right)
    \end{pmatrix}$\\\hline
    QNDI+QNDI & $\begin{pmatrix}
    \eta_B\left(\gamma\cos\epsilon\cos\phi_1-\gamma^{-1}\sin\epsilon\sin\phi_1\right)+\eta_A \cos\phi_2 & -\eta_B \left(\gamma\sin\epsilon\cos\phi_1 + \gamma^{-1}\cos\epsilon\sin\phi_1\right)\\
    -\eta_A \sin\phi_2 & 0
    \end{pmatrix}$\\
       \hline
    \end{tabular}
    \caption{Transmission matrices ${\bf T}_{AB}^{21}$ for different combinations. Here $\theta_{A(B)}$, $r_{A(B)}$ and $\eta_{A(B)}$ are respectively the BS angle, TMS strength and QND strength of the interface $A(B)$.}
    \label{transmission_matrix}
\end{table*}
\end{widetext}

For constructing SWAP, the overall reflection matrix should be engineered to null matrix. We list the explicit expression of the reflection matrix of all combinations on Table.~\ref{reflection_matrix}.
By straight forward calculation, we determine that except for BS+BS combination under the stringent condition, $\chi_A=1-\chi_B$, it is impossible to engineer a null reflection matrix with any $\gamma, \epsilon,\phi_1$ and $\phi_2$. This stringent condition is nothing but the case that the two BS are complementary, i.e. $|\theta_A|+|\theta_B| = \pi/2+n\pi$ for integer $n$, so that the combined interface will be a BS with resultant angle $\theta_{AB}= |\theta_A|+|\theta_B|$, which is essentially a SWAP.

\begin{widetext}
\begin{table*}
    \centering
       \begin{tabular}{|c|c|}
       \hline
Combination & Reflection matrix ${\bf T}_{AB}^{22}$\\
    \hline
    BS+BS   &  $
        \begin{pmatrix}
    \cos\theta_B\cos\theta_A -\sin\theta_B\sin\theta_A \gamma\cos\phi_1 & \sin\theta_B\sin\theta_A \gamma^{-1}\sin\phi_1\\
    -\sin\theta_B\sin\theta_A\gamma\sin\phi_1 & \cos\theta_B \cos\theta_A - \sin\theta_B\sin\theta_A\gamma^{-1}\cos\phi_1
    \end{pmatrix}$\\\hline
    TMS+TMS & $\begin{pmatrix}
    \cosh r_B \cosh r_A +\sinh r_B\cosh r_A\gamma\cos\phi_1 & \sinh r_B\sinh r_A \gamma^{-1}\sin\phi_1\\
    -\sinh r_B\sinh r_A\gamma\sin\phi_1 & \cosh r_B \cosh r_A + \sinh r_B\sinh r_A\gamma^{-1}\cos\phi_1
    \end{pmatrix}$\\\hline
    BS+TMS   & $\begin{pmatrix}
    \cos\theta_B \cosh r_A -\sin\theta_B\sinh r_A\gamma\cos\phi_1 & -\sin\theta_B\sinh r_A \gamma^{-1}\sin\phi_1\\
    -\sin\theta_B\sinh r_A\gamma\sin\phi_1 & \cos\theta_B \cosh r_A + \sin\theta_B\sinh r_A\gamma^{-1}\cos\phi_1
    \end{pmatrix}$\\\hline
    BS+QNDI & $\begin{pmatrix}
\cos\theta_B & -\sin\theta_B \eta_A \left(\gamma \sin\epsilon\cos\phi_1 + \gamma^{-1}\cos\epsilon\sin\phi_1\right)\\
0 & \cos\theta_B - \sin\theta_B\eta_A\left(\gamma \sin\epsilon\sin\phi_1-\gamma^{-1}\cos\epsilon\cos\phi_1\right)\\
    \end{pmatrix}$\\\hline
    TMS+QNDI &  $\begin{pmatrix}
\cosh r_B & \sinh r_B \eta_A\left(\gamma \sin\epsilon\cos\phi_1 + \gamma^{-1}\cos\epsilon\sin\phi_1\right)\\
0 & \cosh r_B - \sinh r_B\eta_A\left(\gamma \sin\epsilon\sin\phi_1-\gamma^{-1}\cos\epsilon\cos\phi_1\right)\\
    \end{pmatrix}$\\\hline
    QNDI+QNDI & $\begin{pmatrix}
    \cos\phi_2 & \sin\phi_2 + \eta_B\eta_A \left(\gamma\sin\epsilon\cos\phi_1+\gamma^{-1}\cos\epsilon\sin\phi_1\right)\\
    -\sin\phi_2 & \cos\phi_2
    \end{pmatrix}$\\
        \hline
    \end{tabular}
    \caption{Reflection matrices ${\bf T}_{AB}^{22}$ for different combinations. }
    \label{reflection_matrix}
\end{table*}
\end{widetext}

\subsection{Supplement for two-interface protocol of BS+BS and TMS+TMS}

In Sec.~\ref{protocolBBTT}, we discussed two special cases. First, when $\chi_A=\chi_B$, we are able to engineer QNDI and Identity. To construct the QNDI, it is required to engineer a rank 1 transmission matrix. According to the results in Table~\ref{transmission_matrix}, we can verify that $\gamma=\tan\phi_1-\sec\phi_1$ and $\phi_1 \neq 0$ is a solution for engineering a QNDI from BS+BS and TMS+TMS combinations. For other combinations of $\gamma$ and $\phi_1$, e.g. the protocol discussed in the beginning of Sec.~\ref{protocolBBTT}, we will obtain a rank 0 transmission matrix and construct the Identity.

Second, when we are considering BS+BS combination and $\chi_A=1-\chi_B$, there are two possible resultant interfaces, sQNDI or SWAP. The requirement for engineering sQNDI is to have a rank 1 reflection matrix. According to the results in Table.~\ref{reflection_matrix}, we can verify that a sQNDI is engineered when $\gamma=-\tan\phi_1-\sec\phi_1$. Any other combinations of $\gamma$ and $\phi_1$ gives SWAP as the resultant interface.

\subsection{Supplement for squeezing-restricted protocol for interfaces with $\chi\neq 0,1$}

In Sec.~\ref{tuning_no_Z}, the squeezing-restricted protocol for interfaces with $\chi\neq 0,1$ requires the identities Eqs.~\eqref{effective_rotation1} and \eqref{effective_rotation2}, i.e.
\begin{eqnarray}
{\bf R}^\alpha_1{\bf \bar{U}}^\chi_{AB} = {\bf R}^{\alpha'}_2 {\bf \bar{U}}^\chi_{AB} {\bf R}^{\alpha''}_1 {\bf R}^{\alpha'''}_2\\
{\bf \bar{U}}^\chi_{CD}{\bf R}^\beta_1 = {\bf R}^{\beta''}_1 {\bf R}^{\beta'''}_2 {\bf \bar{U}}^\chi_{CD} {\bf R}^{\beta'}_2.
\end{eqnarray} 
They are just variants of the interface invariant Eq.~\eqref{exchange}. By using Eqs.~\eqref{exchange}, \eqref{M_resolve}-\eqref{Ly}, we are able to determine the relation between all rotation operations. 
Defining ${\bf R}^{s}_1\equiv{\bf R}_1(\theta_{s})$, ${\bf R}^{s'}_2\equiv{\bf R}_2(\theta_{s}')$, ${\bf R}^{s''}_1\equiv{\bf R}_1(\theta_{s}'')$ and ${\bf R}^{s'''}_2\equiv{\bf R}_2(\theta_{s}''')$ with $s=\alpha, \beta$, the relation between rotation angles are listed in the following table,
\begin{table}[h]
    \centering
    \begin{tabular}{|c|c|c|c|}
    \hline
         Class of Interface        & $\theta_{s}'$ & $\theta_{s}''$ & $\theta_{s}'''$  \\\hline
       $\chi<0$  &  $-\theta_{s}$& $-\theta_{s}$ & $\theta_{s}$\\
        $0<\chi<1$  & $\theta_{s}$ & $-\theta_{s}$ & $\theta_{s}$\\
       $\chi>1$  & $\theta_{s}$ & $\theta_{s}$ & $-\theta_{s}$\\ \hline
    \end{tabular}.
\end{table}

\section{Derivation of pre- and post-squeezing forms}

In this appendix, we will show that a non-trivial interface can always be converted to pre- (Eqs.~(\ref{M_R_chi}), (\ref{M_R_SQ}), (\ref{M_R_Q})) and post-squeezing (Eqs.~(\ref{M_L_chi}), (\ref{M_L_SQ}), (\ref{M_L_Q})) forms under the squeezing restriction by the allowable single-mode controls. 

\subsection{Any non-trivial interface except for QNDI}\label{nonQNDI}

Our aim is to show that there always exists a set of single-mode controls to implement the conversion, i.e. ${\bf L}_1^{\text{out}}{\bf R}^{\text{out}}_{2} {\bf T} {\bf L}_1^{\text{in}}{\bf R}^{\text{in}}_{2} = {\bf \bar{U}}{\bf S}_2$ or ${\bf S}_2{\bf \bar{U}}$. We will show that this identity is satisfied for every block matrix, by illustrating how the suitable single-mode operations are constructed.
First, we can always identify the required mode-2 rotations by the SVD of ${\bf T}^{22}$, i.e.
\begin{eqnarray}\label{T2222}
[{\bf R}^{\text{out}}_{2}]^{22}{\bf T}^{22}[{\bf R}^{\text{in}}_{2}]^{22} = {\bf \Sigma},
\end{eqnarray}
where ${\bf \Sigma} \equiv \text{diag}(\lambda^q_{22},\lambda^p_{22})$ is equal to $[{\bf \bar{U}}{\bf S}_2]^{22}=[{\bf S}_2{\bf \bar{U}}]^{22}$. Explicitly, $\lambda^q_{22}$ and $\lambda^p_{22}$ are given by
\begin{eqnarray}
\lambda^q_{22} &=& \begin{cases}
\Lambda \sqrt{|1-\chi|} & \text{for BS, TMS, sTMS}\\
\Lambda & \text{for sQNDI}
\end{cases},\\
\lambda^p_{22} &=& \begin{cases}
\Lambda^{-1} \sqrt{|1-\chi|} & \text{for BS, TMS}\\
-\Lambda^{-1} \sqrt{|1-\chi|} & \text{for sTMS}\\
0 & \text{for sQNDI}
\end{cases}.
\end{eqnarray}

Next, we can choose ${\bf L}^{\text{in}}_1$ to diagonalize ${\bf T}^{21}$ as following 
\begin{eqnarray}\label{T21}
[{\bf R}^{\text{out}}_2]^{22}{\bf T}^{21} [{\bf L}^{\text{in}}_1]^{11} = {\bf D},
\end{eqnarray}
where ${\bf D} \equiv \text{diag}(\lambda^q_{21},\lambda^p_{21})$ satisfying $ [{\bf \bar{U}} {\bf S}_2]^{21}$ or $[{\bf S}_2{\bf \bar{U}} ]^{21}$ for respectively pre- or post-squeezing form. 
For the post-squeezing form, $\lambda_{21}^q$ and $\lambda_{21}^p$ are given by $\lambda_{21}^q = \Lambda\sqrt{|\chi|}$ and $\lambda_{21}^p = \pm \Lambda^{-1}\sqrt{|\chi|}$, where $+$ is for BS, sTMS and sQNDI, and $-$ is for TMS.
For post-squeezing form, $\lambda_{21}^q = \Lambda\sqrt{|\chi|}$ and $\lambda_{21}^p = \pm \Lambda^{-1}\sqrt{|\chi|}$.

Finally, ${\bf L}_1^{\text{out}}$ is chosen to satisfy the following diagonalization of ${\bf T}^{11}$, 
\begin{eqnarray}\label{T11}
[{\bf L}_1^{\text{out}}]^{11}{\bf T}^{11} [{\bf L}_1^{\text{in}}]^{11} = {\bf \bar{U}}^{11}.
\end{eqnarray}

With the above choices of ${\bf L}_1^{\text{out}}$, ${\bf R}^{\text{out}}_{2} $, ${\bf L}_1^{\text{in}}$ and ${\bf R}^{\text{in}}_{2}$,
${\bf T}^{11}$, ${\bf T}^{21}$, ${\bf T}^{22}$ are transformed into the diagonal form satisfying the pre- or post-squeezing form of ${\bf T}$. Now, we show that ${\bf T}^{12}$ is also transformed to the corresponding diagonal form. First, we consider how the quadratures are transformed under the converted interface: 
\begin{eqnarray}\label{qpqp_qpqp}
\begin{pmatrix}
\hat{q}_1^{\text{out}}\\
\hat{p}_1^{\text{out}}\\
\hat{q}_2^{\text{out}}\\
\hat{p}_2^{\text{out}}
\end{pmatrix}
=
{\bf L}_1^{\text{out}}{\bf R}^{\text{out}}_{2} {\bf T} {\bf L}_1^{\text{in}}{\bf R}^{\text{in}}_{2}
\begin{pmatrix}
\hat{q}_1^{\text{in}}\\
\hat{p}_1^{\text{in}}\\
\hat{q}_2^{\text{in}}\\
\hat{p}_2^{\text{in}}
\end{pmatrix}
\end{eqnarray}
According to Eqs.~\eqref{T2222} and (\ref{T21}), the transformation to mode 2 can be represented as
\begin{eqnarray}
\begin{pmatrix}
\hat{q}_2^{\text{out}}\\
\hat{p}_2^{\text{out}}
\end{pmatrix} 
= {\bf D}
\begin{pmatrix}
\hat{q}_1^{\text{in}}\\
\hat{p}_1^{\text{in}}
\end{pmatrix} 
+ {\bf \Sigma}
\begin{pmatrix}
\hat{q}_2^{\text{in}}\\
\hat{p}_2^{\text{in}}
\end{pmatrix}.
\end{eqnarray}

Then, by the commutation relation between all the output quadratures, we determine that the output mode-1 quadratures must be the linear combination of $(\lambda_{22}^p \hat{q}_1^{\text{in}} - \lambda_{21}^p \hat{q}_2^{\text{in}} )$ and $(\lambda_{22}^q \hat{p}_1^{\text{in}} - \lambda_{21}^q \hat{p}_2^{\text{in}} )$, i.e. 
\begin{eqnarray}\label{qout_pout}
\begin{pmatrix}
\hat{q}_1^{\text{out}}\\
\hat{p}_1^{\text{out}}
\end{pmatrix} 
&=& {\bf V} {\bf \tilde{\Sigma}}
\begin{pmatrix}
\hat{q}_1^{\text{in}}\\
\hat{p}_1^{\text{in}}
\end{pmatrix} 
+
{\bf V}{\bf \tilde{D}}
\begin{pmatrix}
\hat{q}_2^{\text{in}}\\
\hat{p}_2^{\text{in}}
\end{pmatrix},
\end{eqnarray}
where ${\bf \tilde{\Sigma}} \equiv \text{diag}(\lambda_{22}^p,\lambda_{22}^q)$, ${\bf \tilde{D}} \equiv \text{diag}(-\lambda_{21}^p,-\lambda_{21}^q)$ and ${\bf V}$ is a symplectic matrix representing the linear combination. 
Since the transformation ${\bf V}{\bf \tilde{\Sigma}}$, the first term in Eq.~\eqref{qout_pout}, represents the mode-1 reflection, it should be equal to the diagonalized ${\bf T}^{11}$, i.e. ${\bf V}{\bf \tilde{\Sigma}} = {\bf \bar{U}}^{11}$. 
The second term in Eq.~\eqref{qout_pout} represents the mode-1 transmission, and hence ${\bf V \tilde{D}} = [{\bf L}^{\text{out}}_1]^{12} {\bf T}^{12} [{\bf R}^{\text{in}}_2]^{12}$. By using ${\bf V}{\bf \tilde{\Sigma}} = {\bf \bar{U}}^{11}$, we can determine ${\bf V}$ and then obtain 
\begin{eqnarray}\label{vd1}
{\bf V \tilde{D}} = {\bf \bar{U}}^{11} 
\begin{pmatrix}
-\lambda_{21}^p/\lambda_{22}^p & 0 \\
0& -\lambda_{21}^q/\lambda_{22}^q
\end{pmatrix}.
\end{eqnarray} 

Finally, by considering the explicit expression of $\lambda_{22}^q$ and $\lambda_{22}^p$ for each interface, for the pre-squeezing form, Eq.~\eqref{vd1} becomes
\begin{eqnarray}
{\bf V \tilde{D}} =  
\begin{pmatrix}
\Lambda\sqrt{|\chi|} & 0 \\
0& \pm\Lambda \sqrt{|\chi|}
\end{pmatrix}=[{\bf \bar{U}} {\bf S}_2]^{12}.
\end{eqnarray} 
For the post-squeezing form, Eq.~\eqref{vd1} becomes
\begin{eqnarray}
{\bf V \tilde{D}} =  
\begin{pmatrix}
\sqrt{|\chi|} & 0 \\
0& \pm \sqrt{|\chi|}
\end{pmatrix} =[{\bf S}_2{\bf \bar{U}} ]^{12}.
\end{eqnarray} 

\subsection{For QNDI}

For QNDIs, their pre- or post-squeezing form has the non-diagonal mode-2 reflection matrix, so the required single-mode operations are generally different from that determined in the Sec.~\ref{nonQNDI}. We first show that any QNDI can be transformed into the semi-quadrature-diagonal form, i.e. ${\bf L}_1^{\text{out}}{\bf R}^{\text{out}}_{2} {\bf T} {\bf L}_1^{\text{in}}{\bf R}^{\text{in}}_{2} =
{\bf W}$, where
\begin{eqnarray}\label{Wmatrix}
{\bf W}
\equiv \begin{pmatrix}
\Lambda^{-1} & 0 & 0 & 0\\
0  & \Lambda & \kappa \eta \Lambda^2& -\eta \\
\eta & 0 & \Lambda & 0\\
\kappa \eta & 0 & 0 & \Lambda^{-1} \\
\end{pmatrix}.
\end{eqnarray}
Then, we show that ${\bf W}$ can always be converted into the pre- or post-squeezing form by suitable single-mode operations. 

For QNDI, ${\bf T}^{22}$ can still be diagonalized according to Eq.~\eqref{T2222} with $\lambda_{22}^q = (\lambda_{22}^p)^{-1} = \Lambda$. However, Eq.~\eqref{T21} is not guaranteed, since ${\bf T}^{21}$ is of rank 1. 
Instead, we consider $[{\bf R}_2^{\text{out}}]^{22}{\bf T}^{21} \equiv {\bf J}$, where ${\bf J}$ is a general rank-1 matrix and can always be expressed as
\begin{eqnarray}
{\bf J} &=& \begin{pmatrix}
\mu & \alpha \mu \\
\kappa  \mu & \kappa \alpha \mu
\end{pmatrix}.
\end{eqnarray}
To determine the mode-1 operation ${\bf L}_1^{\text{in}}$ for diagonalizing ${
\bf T}^{21}$, we decompose ${\bf J}$ as following
\begin{eqnarray}\label{J_decomposition}
{\bf J} = \begin{pmatrix}
\eta & 0 \\
\kappa \eta & 0
\end{pmatrix} [{\bf S}_1\left(\frac{\mu \sqrt{1+\alpha^2}}{\eta}\right) {\bf R}_1(\phi_{LQ})]^{11},
\end{eqnarray}
where $\tan\phi_{LQ}\equiv -\alpha$. In Eq.~\eqref{J_decomposition}, the rotation ${\bf R}_1(\phi_{LQ})$ is determined from the LQ decomposition \cite{qu_decomp} of ${\bf J}$, and the squeezing ${\bf S}_{1}(\mu\sqrt{1+\alpha^2}/\eta)$ is set such that the QND strength of ${\bf T}$, denoted by the upper diagonal element of leftmost matrix on the right hand side of Eq.~\eqref{J_decomposition}, is manipulated from $\mu\sqrt{1+\alpha^2}$ to $\eta$.
Comparing to Eq.~\eqref{N_eq}, we determine ${\bf L}_1^{\text{in}} = \left[{\bf S}_1\left(\mu \sqrt{1+\alpha^2}/\eta\right) {\bf R}_1(\phi_{LQ})\right]^{-1}$ and 
\begin{eqnarray}\label{QNDI_T21}
[{\bf R}_2^{\text{out}}]^{22}{\bf T}^{21} [{\bf L}_1^{\text{in}}]^{11} =  \begin{pmatrix}
\eta & 0 \\
\kappa \eta & 0
\end{pmatrix}.
\end{eqnarray}
Then, ${\bf L}_1^{\text{out}}$ is chosen to satisfy
\begin{eqnarray}\label{QNDI_T11}
[{\bf L}_{1}^{\text{out}}]^{11} {\bf T}^{11} [{\bf L}_1^{\text{in}}]^{11} = 
\begin{pmatrix}
\Lambda^{-1} & 0 \\
0 & \Lambda
\end{pmatrix}.
\end{eqnarray}

Considering the transformation of the quadratures under the converted interface, Eq.~\eqref{qpqp_qpqp}, and its expression Eqs.~(\ref{T2222}), (\ref{QNDI_T21}) and (\ref{QNDI_T11}), the output transformed mode-2 quadratures should be given by
\begin{eqnarray}
\begin{pmatrix}
\hat{q}_2^{\text{out}}\\
\hat{p}_2^{\text{out}}
\end{pmatrix} 
= \begin{pmatrix}
\eta & 0 \\
\kappa \eta & 0
\end{pmatrix}
\begin{pmatrix}
\hat{q}_1^{\text{in}}\\
\hat{p}_1^{\text{in}}
\end{pmatrix} 
+ \begin{pmatrix}
\Lambda & 0 \\
0 & \Lambda^{-1}
\end{pmatrix}
\begin{pmatrix}
\hat{q}_2^{\text{in}}\\
\hat{p}_2^{\text{in}}
\end{pmatrix},\nonumber
\end{eqnarray}
and the output transformed mode-1 quadratures should be written as
\begin{eqnarray}
\begin{pmatrix}
\hat{q}_1^{\text{out}}\\
\hat{p}_1^{\text{out}}
\end{pmatrix} 
= \begin{pmatrix}
\Lambda^{-1} & 0 \\
0 & \Lambda
\end{pmatrix}
\begin{pmatrix}
\hat{q}_1^{\text{in}}\\
\hat{p}_1^{\text{in}}
\end{pmatrix} 
+ {\bf M}
\begin{pmatrix}
\hat{q}_2^{\text{in}}\\
\hat{p}_2^{\text{in}}
\end{pmatrix}.
\end{eqnarray}
Here ${\bf M}$ is determined by the commutation relation of the output quadratures, and it is equal to
\begin{eqnarray}
{\bf M}  =
\begin{pmatrix}
0 & 0 \\
\kappa \eta \Lambda^2 & -\eta
\end{pmatrix}.
\end{eqnarray}
These results imply that any QNDI can be transformed into the semi-quadrature-diagonal form ${\bf W}$ (Eq.~\eqref{Wmatrix}).

Finally, by direct calculation, we show that ${\bf W}$ can be rewritten as
\begin{eqnarray}
{\bf W} = {\bf S}_1(\Lambda_R){\bf R}_2(-\phi_R) [{\bf \bar{U}}^Q(\eta) {\bf R}_2(\phi_R){\bf S}_2(\Lambda)]{\bf S}_1(\Lambda_R'),\nonumber\\
\end{eqnarray}
where $\tan\phi_R \equiv \kappa$ and $\Lambda_R \equiv (\Lambda\sqrt{1+\kappa^2})^{-1}$, $\Lambda_R' \equiv \sqrt{1+\kappa^2}$. Therefore, the pre-squeezing form can be achieved by undoing the squeezing ${\bf S}_1(\Lambda_R)$ and rotation ${\bf R}_2(-\phi_R)$ after the interface, and  ${\bf S}_1(\Lambda_R')$ before the interface.

For post-squeezing form, we rewrite ${\bf W}$ as
\begin{eqnarray}
{\bf W} ={\bf S}_1(\Lambda_L)[{\bf S}_2(\Lambda){\bf R}_2(\phi_L)  {\bf \bar{U}}^Q(\eta)] {\bf S}_1(\Lambda_L') {\bf R}_2(-\phi_L),\nonumber\\
\end{eqnarray}
where $\tan\phi_L \equiv -\kappa\Lambda^2$, $\Lambda_L \equiv(\Lambda\sqrt{1+\kappa^2\Lambda^4})^{-1}$ and $\Lambda_L' \equiv \sqrt{1+\kappa^2\Lambda^4}\Lambda^{-1}$. Similarly, the post-squeezing form is achieved after undoing the squeezing ${\bf S}_1(\Lambda_L')$ and rotation ${\bf R}_2(-\phi_L)$ before the interface, and ${\bf S}_1(\Lambda_L)$ before the interface.

\section{Details of irreducible squeezing}

\subsection{Engineering interfaces with $\chi\neq 0,1$}

In the protocol engineering interfaces with $\chi\neq 0,1$ in Sec.~\ref{sec_irreducible_squeezing}, two intermediate interfaces $AB$ and $CD$ are cascaded according to 
\begin{eqnarray}\label{Txi}
{\bf T}_\zeta \equiv {\bf \bar{U}}^\chi_{CD} {\bf S}_1(\gamma) {\bf F}_1{\bf \bar{U}}^\chi_{AB},
\end{eqnarray}
where ${\bf T}_\zeta$ denotes an interface with $\chi \neq 0,1$. Under this configuration, the overall transmission strength, $\chi_{ABCD}\equiv\text{det}[{\bf T}_\zeta]$, is independent of the mode-1 squeezing $\gamma$, but the resultant irreducible squeezing is manipulated by $\gamma$. 
In this appendix, we investigate the irreducible squeezing of ${\bf T}_\zeta$.

By directly calculating the singular values of ${\bf T}_\zeta^{22}$, we determine irreducible squeezing as 
\begin{eqnarray}
\Lambda_{a} = \sqrt{|X|+\Gamma+\sqrt{\Gamma^2+2|X|\Gamma}},
\end{eqnarray}
where
\begin{eqnarray}
&& X =\frac{1-\chi_{ABCD}}{|1-\chi_{ABCD}|},
\end{eqnarray}
and the expression of $\Gamma \equiv \Gamma(\gamma)$ is different for each combination. Only four combinations of intermediate interfaces give unbounded irreducible squeezing, they are BS+BS, TMS+TMS, TMS+sTMS and sTMS+sTMS. It is because their $\Gamma$ are given by
\begin{eqnarray}
\Gamma = \frac{|\chi_{AB}\chi_{CD}|}{2|1-\chi_{ABCD}|}\left(\frac{\gamma^2-1}{\gamma}\right)^2, \label{Gamma_BBTT}
\end{eqnarray}
that can be tuned to arbitrary positive value by choosing the suitable single-mode squeezing $\gamma$. This allows us to engineer $\Lambda_a$ to be any value larger than or equal to $1$. It is straightforward to show that $\Lambda_a \geq \sqrt{|X|}=1$ because $\Gamma \geq 0$. Therefore, the resultant irreducible squeezing can be tuned to any desired value according to Eqs.~\eqref{lambdaABCD} and \eqref{lambdaABCD2}. 

We note that our protocol in Sec.~\ref{sec_irreducible_squeezing} is designed to avoid other combinations since they give the bounded irreducible squeezing or cannot engineer the desired transmission strength. For BS+TMS and BS+sTMS, their $\Gamma$ is given by
\begin{eqnarray}
\Gamma = \frac{|\chi_{AB}\chi_{CD}|}{2|1-\chi_{ABCD}|}\left(\frac{\gamma^2+1}{\gamma}\right)^2, \label{Gamma_BT}
\end{eqnarray}
that has the lower bound $2|\chi_{AB}\chi_{CD}|/|1-\chi_{ABCD}|$, and hence $\Lambda_a$ is bounded. For the combinations involving one QND or sQND, we lose the ability to introduce the controllable rotation ${\bf R}_2^{\beta'''}$ to quadrature-diagonalize ${\bf T}_\zeta$, since Eq.~\eqref{effective_rotation2} is invalid for QNDI or sQND due to the constraint Eq.~\eqref{LpQND}, which tells only shearing operation can satisfy Eq.~\eqref{exchange}. Without the ability to quadrature-diagonalize ${\bf T}_\zeta$, 
the resultant irreducible squeezing may be bounded. Finally, for QNDI+QNDI, QNDI+sQNDI and sQNDI+sQNDI, they must generate an interface with $\chi=0$ or $1$ according to Eq.~\eqref{Txi} that is not the aim of the protocol. 

\subsection{Engineering sQNDI}

Second, we provide the detail calculation for the resultant irreducible squeezing for the sQNDI protocol in Sec.~\ref{SQND}. After cascading the interfaces $AB$ and $CD$ according to Fig.~\ref{fig_SQND}(a), we have the configuration Fig.~\ref{fig_SQND}(b),
\begin{eqnarray}
{\bf T}_{ABCD} \equiv {\bf S}_2(\Lambda_{CD}){\bf R}_2(\phi_{CD}) {\bf \bar{U}}^Q(\gamma \eta_{CD}+\Lambda_{AB}) {\bf \bar{U}}^S,\nonumber\\
\end{eqnarray}
and then the mode-2 reflection matrix is given by
\begin{eqnarray}
{\bf T}_{ABCD}^{22} = [{\bf S}_2(\Lambda_{CD}){\bf R}_2(\phi_{CD})]^{22} \begin{pmatrix}
\gamma \eta_{CD}+\Lambda_{AB} & 0\\
0 & 0
\end{pmatrix}.\nonumber\\
\end{eqnarray}
By direct calculation, the non-zero singular value is given by
\begin{eqnarray}
\left(\gamma \eta_{CD} +\Lambda_{AB}\right)
\frac{\sqrt{(\Lambda_{CD}^4+1)+(\Lambda_{CD}^4-1)\cos\phi_{CD}}}{2\Lambda_{CD}},\nonumber\\
\end{eqnarray}
This is the QND strength, and hence the irreducible squeezing strength, of the resultant sQNDI (c.f. Eq.~\eqref{sigma_sq}).

\section{Alternative remote squeezing protocols}

In this appendix, we present two alternative remote squeezing schemes that uses fewer component interfaces than the standard scheme in Sec.~\ref{remote_gate} but works under specific conditions. 

The first alternative scheme applies the same idea of the standard scheme that requires $N$ component interfaces in total and uses $N-1$ component interfaces to engineer an intermediate interface that is the inverse of the last component interface preceded with the desired mode-2 squeezing. The standard scheme requires $N=5$ since the intermediate interface is engineered according to the four-interface protocols in Sec.~\ref{sec_irreducible_squeezing}-\ref{QND}. We note that it is possible to engineer the intermediate interface with fewer component interfaces if we can prepare particular components. To engineer the intermediate interface with $\chi \neq 0,1$, we can use three component interfaces by replacing the interface $AB$ with the component interface $A$ (or $B$) in the protocol in Sec.~\ref{sec_irreducible_squeezing}, if $A$ (or $B$) satisfies the criteria Eq.~\eqref{condition_for_chiAB}. Then, we can have a four-interface protocol for remote squeezing. To engineer a sQNDI as the intermediate interface, it is possible to use only 2 components if one component is QNDI and another component is sQNDI. Then we can directly cascade them according to Fig.~\ref{fig_SQND}(b). Similar for engineering a QNDI as the intermediate interface, it requires only 2 component interfaces if we can prepare two sQNDIs as components.

The second alternative scheme requires four non-trivial component interfaces $A$, $B$, $C$ and $D$ that are neither QNDI nor sQNDI. The strategy is to engineer two intermediate interfaces $AB$ and $CD$ with the same $\chi$ such that $AB$ and $CD$ are inverse of each other up to the mode-2 operation. 
The explicit protocol is
\begin{enumerate}
  \item Converting $A$ ($C$) and $B$ ($D$) into respectively the pre- and post-squeezing forms, same as the first step of the two-interface module in Sec.~\ref{two_interface_module};
  \item Combining interfaces $A$ with $B$ and $C$ with $D$ to form respectively the intermediate interface $AB$ and $CD$ with the same transmission strength, i.e. $\chi_{AB}=\chi_{CD}\equiv\chi_{\text{int}}$ according to
  \begin{eqnarray}\label{tabcd}
  {\bf T}_{AB(CD)} \equiv {\bf S}_{2}(\Lambda_{B(D)})
 {\bf \bar{U}}^\chi_{B(D)} {\bf S}_1 {\bf \bar{U}}^\chi_{A(C)} {\bf S}_2(\Lambda_{A(C)});\nonumber\\
  \end{eqnarray}
 \item Applying suitable single-mode controls on $AB$ according to Eqs.~\eqref{U_chi}-\eqref{U_SQ}, such that $AB$ is engineered to be the inverse of $CD$ with the specific mode-2 squeezing, i.e. ${\bf T}_{AB}' = {\bf T}^{-1}_{CD} {\bf S}_{2}(\Lambda_{ABCD})$, where $\Lambda_{ABCD}=\Lambda_{AB}\Lambda_{CD}$;
 \item Tuning $\chi_{\text{int}}$ such that the resultant remote squeezing $\Lambda_{ABCD}=\Lambda_{\text{tgt}}$.
\end{enumerate}
According to Eq.~\eqref{tabcd}, the irreducible squeezing of the intermediate interface is given by
\begin{eqnarray}
\Lambda_{AB(CD)} &=& \frac{\Lambda_{A(C)}\Lambda_{B(D)}}{\sqrt{|1-\chi_{\text{int}}|}}\Big[ 2-\chi_{A(C)}-\chi_{B(D)}-\chi_{\text{int}}\nonumber\\
&+ &\sqrt{(\chi_{\text{int}}-X_{f,+})(\chi_{\text{int}}-X_{f,-})}\Big],
\end{eqnarray}
where $X_{f,\pm} \equiv  \pm 2 \sqrt{\chi_{A(C)}\chi_{B(D)}(1-\chi_{A(C)})(1-\chi_{B(D)})}+\chi_{A(C)}+\chi_{B(D)}-2\chi_{A(C)}\chi_{B(D)}$. We note that $\Lambda_{AB(CD)}$ has no bound and hence $\Lambda_{ABCD}$ can be tuned to arbitrary values. 

To summarize, these two alternative protocols requires at most four component interfaces, but they have limitations, both of them requires particular component interfaces. The first protocol considers the special case of the protocols in Sec.~\ref{sec_irreducible_squeezing}-\ref{QND}, and the second protocol works when the components are BS, TMS or sTMS. If we have ability to choose the type of component interfaces in a platform, these alternative protocols are preferred since they requires fewer number of interfacing. It may reduce the implementation time and the chance of the operation error.


\bibliography{apsformat}
\pagestyle{plain}

\end{document}